\documentclass[12pt,reqno]{amsart}
\usepackage{amsmath,amsthm,bbold}

\theoremstyle{plain}

\theoremstyle{definition}

\theoremstyle{remark}

\numberwithin{equation}{section}
\numberwithin{theorem}{section}
\numberwithin{figure}{section}
\numberwithin{table}{section}

\newcommand{\bK}{{\mathbf K}} \newcommand{\bz}{{\mathbf z}} 
 
\newcommand{\bm}{{\mathbf m}}  
\newcommand{\bfe}{{\mathbf e}}
\newcommand{\bu}{{\mathbf u}} \newcommand{\bp}{{\mathbf p}} 
\newcommand{\bt}{{\mathbf t}} 
  
\newcommand{\bQ}{{\mathbf Q}} \newcommand{\bL}{{\mathbf L}} 

\newcommand{\bs}{\mathbf s}

\newcommand{\bB}{\mathbf B}
\newcommand{\bA}{\mathbf A}
\newcommand{\bD}{\mathbf D}
\newcommand{\bj}{\mathbf j}
\newcommand{\bl}{\mathbf l}

\newcommand{\cC}{\mathcal{C}}
\newcommand{\cS}{\mathcal{S}}

\newfont{\BBB}{msbm10 scaled\magstep1}
\newcommand{\ZZ}{\mbox{\BBB Z}}

\newcommand{\id}{{\mathbb 1}}

\newcommand{\del}{\partial}

\newcommand{\la}{\lambda}

\newcommand{\latot}{\lambda_{\text{tot}}}
\newcommand{\mutot}{\mu_{\text{tot}}}

\DeclareMathOperator{\tr}{Tr}
\DeclareMathOperator{\charac}{ch}

\newcommand{\qbin}[2]{\genfrac{[}{]}{0pt}{}{#1}{#2}}

\newcommand{\wh}[1]{\widehat{#1}}
\newcommand{\eql}{=} 
\newcommand{\qn}[1]{(q)_{#1}}   

\newcommand{\ua}{\uparrow}     \newcommand{\da}{\downarrow}

\newcommand{\txt}[1]{{\textstyle{#1}}}
\newcommand{\scr}[1]{{\scriptstyle{#1}}}

\begin{document}


\title[Non-abelian quantum Hall states]{Non-abelian quantum Hall states \\
-- exclusion statistics, $K$-matrices and duality --}


\author[E.~Ardonne]{Eddy Ardonne}
\address{Institute for Theoretical Physics, University of Amsterdam,
Valckenierstraat 65, 1018~XE~~Amsterdam, The Netherlands}
\email{ardonne@wins.uva.nl}
\author[P.~Bouwknegt]{Peter Bouwknegt}
\address{Department of Physics and Mathematical Physics, University 
of Adelaide, Adelaide SA~5005, Australia}
\email{pbouwkne@physics.adelaide.edu.au}
\author[K.~Schoutens]{Kareljan Schoutens}
\address{Institute for Theoretical Physics, University of Amsterdam,
Valckenierstraat 65, 1018~XE~~Amsterdam, The Netherlands}
\email{kjs@wins.uva.nl}


\dedicatory{Dedicated to Rodney J.~Baxter on his 60th birthday}
\thanks{ITFA-00-07, ADP-99-23/M82}



\begin{abstract}
We study excitations in edge theories for non-abelian
quantum Hall states, focussing on the spin polarized
states proposed by Read and Rezayi and on the spin singlet 
states proposed by two of the authors. By studying the 
exclusion statistics properties of edge-electrons and 
edge-quasiholes, we arrive at a novel $K$-matrix structure.
Interestingly, the duality between the electron and 
quasihole sectors links the pseudoparticles that are 
characteristic for non-abelian statistics with composite
particles that are associated to the `pairing physics'
of the non-abelian quantum Hall states.
\end{abstract}


\maketitle

\section{Introduction}

The fractional quantum Hall effect has 
led to the identification of new states of matter, which
can be characterized as incompressible quantum fluids
with off-diagonal long-range order (`topological order').
After the initial discovery of the `principal Laughlin
series' of quantum Hall fluids at filling factor
$\nu=1/m$, a large class of so-called abelian 
quantum Hall fluids has been identified, accounting for 
the rich spectrum of fractional quantum Hall plateaus
that have been observed in the lowest Landau level. 

The observation of a quantum Hall plateau at filling 
factor $\nu=5/2$ (see \cite{Panetal} for a recent experiment)
has made it clear that the 
traditional set of abelian quantum Hall states (which all
share the property of having an odd-denominator filling
factor) will not suffice for explaining the phenomena
observed in the second Landau level. Prompted by this 
development, new categories of incompressible quantum
fluids have been proposed. Among them are various `paired'
or `clustered' states, such as the Pfaffian states first 
proposed by Moore and Read \cite{MR}. The quasiparticles over 
these states satisfy what is called non-abelian braid 
statistics, and by abuse of language one speaks of 
`non-abelian quantum Hall states'.

The characteristic order of the abelian quantum Hall fluids 
should be viewed as `topological' and it can be 
characterized by a collection of integer numbers, which together 
constitute a so-called $K$-matrix. Many of the low energy 
characteristics of the Hall fluid are encoded in this
$K$-matrix and the 
quantum numbers of the elementary electron-type excitations.
They include the filling factor $\nu$ and the spin Hall
conductance $\sigma$. In addition the (fractional) quantum
numbers of the various quasihole type excitations are
determined by using (the inverse of) the $K$-matrix.

A systematic framework for the physical implications of the
topological order embodied in the $K$-matrix is provided
by effective Chern-Simons and conformal field theories for
bulk and edge excitations, respectively. In a systematic 
treatment of the low energy dynamics, these theories arise as 
special limits of a unifying field theory for the low energy 
behaviour of quantum Hall systems \cite{PSB}.

It is well-known that bulk-excitations over fractional quantum
Hall fluids satisfy fractional (anyonic) braid statistics.
Closely related to this are the fractional exclusion statistics
of both bulk \cite{ICJ} and edge excitations \cite{WY,ES}. 
It has been observed \cite{FK,ES} 
that for abelian quantum Hall fluids, the (edge) statistics 
matrix $\bK$ (in the sense of Haldane's definition of exclusion 
statistics \cite{Halb}) is closely related to the $K$-matrix.

The main purpose of the present paper is to present a $K$-matrix
structure associated to specific series of non-abelian quantum
Hall states. To this end, we study the exclusion statistics of
edge excitations over these quantum Hall states, and identify
from that analysis statistics matrices $\bK$. We shall then 
argue that these same matrices can be viewed as $K$-matrices
for these non-abelian quantum Hall states. 

Our analysis here builds on earlier results published in 
\cite{Scb,AS,GS,BCR,BSb}. In \cite{ABGS} we presented our
present results in brief form, and elaborated on the
physical meaning of the newly obtained $K$-matrices.

This paper is organized
as follows. In Sect.\ 2 we briefly review the $K$-matrix 
theory for abelian quantum Hall states, and make a generalization
in order to be able to treat spin singlet states. We continue
in Sect.\ 3 by making the link to abelian exclusion statistics. 
We argue that the statistics matrix is (basically) given 
by the $K$-matrix. Also, we introduce an important notion
of duality. In Sect.\ 4 we  generalize this concept to 
the non-abelian case, where composites and pseudoparticles 
play a vital role. It is argued that the well known 
formulas for the physical quantities such as the 
filling factor, derived from the abelian $K$-matrix structure,
still hold for the non-abelian $K$-matrices describing
various clustered non-abelian quantum Hall states.   
Sect.\ 5 deals with the relation between the 
universal chiral partition function (UCPF) and exclusion
statistics. In Sect.\ 6 and 7 the $K$-matrices for two
classes of non-abelian clustered states are identified.
Sect.\ 8 is reserved for discussions, while some of the
more mathematical results, for instance on character 
formulas, are discussed in the appendices.

\section{$K$-matrices for abelian quantum Hall states}

In this section, we briefly review the $K$-matrix
structure for abelian quantum Hall states. We do not 
derive, but merely state the results we need in this paper. 
For a more detailed review, see for instance \cite{Wenb}.

The information needed to describe an abelian quantum 
Hall state can be encoded in the following way,
henceforth referred to as the fqH data. The four  
important `objects' which will do the job 
are the $K$-matrix (which will also play the role as 
statistics matrix), the so called  charge and spin vectors,
$\bt$ and $\bs$, respectively, and the angular momentum vector $\bj$.
A few remarks with respect to the notation of spin vectors
need to be made at this point. In \cite{WZa,WZb}, the concept of a 
`spin vector' was introduced. This `spin vector'
is in fact related to the angular momentum
of the electrons on (for instance) the sphere and is needed to 
calculate the so-called shift. 
In our case we need to distinguish between this angular momentum
vector and the vector containing the real spin of the particles.
Therefore, we have denoted the angular momentum vector by $\bj$, 
and the vector containing the spin quantum numbers by $\bs$. 

In order to have the possibility to connect the $K$-matrix with
the statistics matrix (as we will do in the following sections),
we will distinguish between the $K$-matrix for
the `electron part' and the `quasihole part' of the theory.
These will be denoted by $\bK_e$ and $\bK_\phi$, respectively.
The corresponding charge, spin and angular momentum vectors are 
$\bt_e$, $\bt_\phi$, $\bs_e$, $\bs_\phi$, $\bj_e$, and $\bj_\phi$ 
in an obvious notation.
In all the cases we considered, it is possible to choose 
a basis in which the $K$-matrices are just each others inverse,
$\bK_e = \bK_\phi^{-1}$.

As stated above, the $K$-matrices will play several roles in the 
theory. First of all, they couple the different Chern-Simons gauge 
fields which play a central role in a Lagrangian description 
of the quantum Hall states. In the abelian case, the Chern-Simons
part of the Lagrangian for a system on a surface of genus $g$
reads as follows
\begin{equation} \label{cslag}
{\mathcal L}_{\text{CS}} \eql 
\frac{1}{4\pi}  \epsilon^{\mu \nu \lambda} \left(
{\bf K}^{ij}_e a^{i}_{\mu} \del_{\nu} a^{j}_{\lambda} 
+ 2 \bt_e^i A_\mu \del_\nu a^i_\lambda 
+ 2 \bj_e^i \omega_\mu \del_\nu a^i_\lambda
+ 2 \bs_e^i \beta_\mu \del_\nu a^i_\lambda \right) \,,
\end{equation}
where the fields $a$ are the Chern-Simons gauge fields. The Greek 
indices run over $\{0,1,2\}$, and the Roman indices over
the number of channels. 
The first three terms in Eq.\ \eqref{cslag} are rather standard
and described in, for instance, \cite{Wenb,WZa,WZb}. The first term is
the famous Chern-Simons term, the other three describe the
couplings to various fields. The gauge field $A_\mu$ describes the
electromagnetic field and $\omega_\mu$ is the `spin connection' which 
gives rise to the curvature of the space on which the system is defined. 
To explain the last term, we briefly discuss the concept of the
spin Hall conductance and the related spin filling factor
$\sigma$ (see \cite{RG} and references therein). 

In general, one would define the spin conductance in the same
way as the charge conductance, namely as a response to a
certain field. In the case of a quantum Hall system, the role
of the electric field is taken over by a gradient in the Zeeman
energy. The gauge field describing this is denoted by $\beta_\mu$ in
Eq.\ \eqref{cslag}. The spin Hall conductance is then related
to the `spin-current' induced perpendicular to the direction
of the gradient of the Zeeman energy. 

Let us now briefly recall the results obtained from this
formulation for the filling factors and the shift 
corresponding to a surface of genus $g$. 
The filling factors can be calculated by means of simple inner products%
\footnote{Throughout this paper the transpose in equations like 
\eqref{sec2eq2ea} is implicitly understood in order to simplify the 
notation.}
\begin{eqnarray} \label{sec2eq2ea} 
&& \nu \eql \bt_e \cdot \bK^{-1}_e \cdot \bt_e \eql  
\bt_{\phi} \cdot \bK^{-1}_{\phi} \cdot \bt_{\phi}\,,  \nonumber \\
&& \sigma  \eql  \bs_e \cdot \bK^{-1}_e \cdot \bs_e \eql 
\bs_{\phi} \cdot \bK^{-1}_{\phi} \cdot \bs_{\phi} \,.
\end{eqnarray}
The relation between the charge (and spin) vectors of the electron and
quasihole parts are given by
\begin{equation} \label{sec2eq3ea}
\bt_\phi  \eql  - \bK^{-1}_e \cdot \bt_e\,, \qquad
\bs_\phi  \eql  - \bK^{-1}_e \cdot \bs_e \,.
\end{equation}
The last important property we will discuss is the so called `shift'
in the flux on surfaces of general genus $g$. The relation
between the number of electrons $N_e$ and the corresponding 
number of flux quanta $N_\Phi$ is given by
\begin{equation} \label{sec2eq4ea}
N_\Phi \eql  \frac{1}{\nu} N_e - \mathcal{S} \,,
\end{equation}
where the shift $\mathcal{S}$ is given by
\begin{equation} \label{sec2eq5ea}
\mathcal{S} \eql \frac{2(1-g)}{\nu}\, (\bt_e \cdot \bK_e^{-1} \cdot 
\bj_e) \,.
\end{equation}
Although $\bj_e$ plays a somewhat different role than $\bt_e$
and  $\bs_e$, we define $\bj_\phi$ by analogy to \eqref{sec2eq3ea}
\begin{equation}
\bj_\phi \eql  - \bK^{-1}_e \cdot \bj_e\,.
\end{equation}
In the present paper, we shall establish that the various relations 
given above are not just valid for the abelian case. They also apply 
in the non-abelian case,
under the condition that a formulation is used in which
the pseudoparticles do not carry charge or spin (see Sect.\ 4).
We shall see that in all the cases we consider, such a formulation
can indeed be given.

The other important role the $K$-matrices play will be
described in the next section, namely the role as statistics matrix
in the sense of the Haldane exclusion statistics of the 
(quasi) particles. Also, we will explain a notion
of `duality' which is important in this context, and rederive
some of the relations given above.

\section{Abelian exclusion statistics}

An important consequence of the concept of an
`ideal gas of fractional statistics particles' is the notion
of 1-particle distribution functions which generalize the
familiar Fermi-Dirac and Bose-Einstein distributions.
These distributions can be derived from `1-particle grand 
canonical partition functions'. These quantities, which
we denote by $\lambda_i$, satisfy the following 
set of equations, which were independently derived by Isakov, 
Dasni\`eres de Veigy-Ouvry and Wu (IOW) \cite{IOW}
\begin{equation} \label{sec3eq1ea}
\left( \frac{\lambda_i-1}{\lambda_i} \right) \prod_j 
\lambda_j^{\bK_{ij}} = z_i \,,
\end{equation}
where $\lambda_i = \lambda_i (z_1,\ldots,z_n)$,
with $z_i=e^{\beta(\mu_i-\epsilon)}$ the generalized fugacity 
of species $i$.  Note that the energy $\epsilon$ may also 
include contributions from the coupling of the charge and spin
of the quasiparticles to external electric and magnetic fields.
Hence the information about 
charge and spin of the quasiparticles is also encoded in these 
generalized fugacities.
The fugacities of the particles will be important for the
distinction between abelian and non-abelian statistics, as 
we will point out later. The matrix $\bK$ is the so-called 
`statistics matrix' and describes, at least in 
the original situation in which Haldane introduced his new
notion of statistics, the statistical interaction of 
particles of different species. 

{}From the solutions $\lambda_i$ of the IOW-equations
\eqref{sec3eq1ea} the one-particle distribution
functions $n_i(\epsilon)$ are obtained as
\begin{equation} \label{sec3eq2ea}
n_i(\epsilon) \eql 
z_i \frac{\partial}{\partial z_i} \log \prod_j \lambda_j 
{}_{\big| z_i=e^{\beta(\mu_i-\epsilon)}} \eql
\sum_j z_j \frac{\partial}{\partial z_j} \log \lambda_i
{}_{\big| z_i=e^{\beta(\mu_i-\epsilon)}} \ ,
\end{equation}
where we have assumed that the matrix $\bK$ is symmetric.

The relation between, on the one hand, the $K$-matrix of 
an abelian quantum Hall fluid and, on the other hand, the
exclusion statistics of its charged edge excitations,
can be described as follows. The charged edge excitations
are described by a specific Conformal Field Theory (CFT),
also known as a chiral Luttinger liquid. Following
a procedure first proposed in \cite{Sca}, one may
associate a notion of fractional exclusion statistics
to a set of fundamental excitations in this CFT.
Selecting a particular set of negatively charged
`electron type' excitations together with a
`dual' set of positively charged quasihole
excitations, one precisely finds fractional exclusion 
statistics in the sense of Haldane, with statistics matrix 
$\bK$ given by
\begin{equation} \label{sec3eq3ea}
\bK \eql \bK_e \oplus \bK_\phi\, ,
\end{equation} 
with $\bK_e$ and $\bK_\phi$ the $K$-matrices for
the abelian quantum Hall state. For the principal
Laughlin series at filling fraction $\nu=1/m$,
this result was obtained in \cite{ES}, in 
its general form it first appeared in our paper
\cite{ABGS}. The relation of the identification
\eqref{sec3eq3ea} with character identities involving
so called Universal Chiral Partition Functions
will be discussed in Sect.\ 5.

In \cite{FK}, a slightly different identification
between the $K$-matrix and a statistics matrix, 
amounting to 
$\bK=\bK_e$, was proposed. The two proposals can 
be reconciled by realizing that we, in our analysis
of edge excitations, restrict ourselves to
quanta of positive energy only. From the duality
relations that we discuss below, one learns that,
in a precise sense, quasihole quanta of
positive energy can be traded for holes in a `Fermi sea' 
of electron-type quanta at negative energy, and
in this way one arrives at a complete description in 
terms of the matrix $\bK_e$ alone.

One of the main themes in this paper will be the 
identification of statistics matrices $\bK$ for 
excitations over non-abelian quantum Hall states.
Extending the identification \eqref{sec3eq3ea}
to the non-abelian case, we shall propose
$K$-matrices for the non-abelian quantum Hall
states. We would like to stress that, although many 
of the formulas from the well known abelian 
$K$-matrix description still hold for the generalized 
$K$-matrices we find here, the description for the
non-abelian states is on an entirely 
different footing. The abelian $K$-matrices were introduced 
to describe quantum Hall states in the `most general' way, i.e.\ 
by trying to implement the hierarchical schemes in a general way. 
In the non-abelian case, we need the $K$-matrix structure to 
keep track of the non-abelian statistics. So although we use 
a matrix structure, we are not describing a hierarchical situation.    

We continue this section with a discussion of the fundamental
`particle-hole' duality between the electron and the 
quasihole sectors of the theory. To show how this duality 
works, we assume that we have $n$ quasiholes $\phi$ and $n$ 
electron-like particles $\Psi$ described by the matrices $\bK_\phi$ 
and $\bK_e$, respectively. We assume that 
(i) $\bK_\phi = \bK_e^{-1}$, and
(ii) there is no mutual exclusion statistics between the two
sectors (meaning that the statistics matrix is given by the 
direct sum \eqref{sec3eq3ea}). These two conditions in fact 
constitute what we mean by duality in this context.
In the context of low-energy effective actions for abelian fqH
systems, a similar notion of duality has been considered
(see, e.g. \cite{WZb} and references therein).

With the matrices $\bK_\phi$ and $\bK_e$, two independent
systems of IOW-equations can be written down, and these systems
are related by the duality (for clarity, we will denote the single
level partition function for the quasiholes and
electron-like particles  
by $\lambda_i$ and $\mu_i$ respectively; the corresponding 
fugacities will be denoted by $x_i$ and $y_i$)
\begin{equation} \label{sec3eq4ea}
\lambda_i \eql  \frac{\mu_i}{\mu_i-1} \,, \qquad 
x_i \eql \prod_j y_j^{-(\bK_e)_{ij}^{-1}} \,,
\end{equation}
as can be verified easily. 

As an illustration of the duality, we calculate
the central charge of the conformal field theory that
describes the edge excitations. We focus on the abelian case. 
In the non-abelian case, which we discuss in the next section, 
there will be a subtraction term due to the presence of 
pseudoparticles.
 
In general, for abelian quantum Hall states,
the central charge $c_{\text{CFT}}$ is given by 
\begin{equation} \label{sec3eq10ea}
c_{\text{CFT}} \eql {\frac{6}{\pi^2}}
\int_0^1 \frac{dz}{z} \log \latot(z) \ ,
\end{equation}
where $\latot(z)$ denotes the product $\prod_j \lambda_j$
evaluated at $z_j=z$ for all $j$. It has been shown 
(see \cite{BSb,BCR} and references therein),
that this can be rewritten in the following form
\begin{equation} \label{sec3eq11ea}
c_{\text{CFT}} \eql  {\frac{6}{\pi^2}} \sum_i L(\xi_i) \,,
\end{equation}
where $L(z)$ is Rogers' dilogarithm
\begin{equation} \label{sec3eq12ea}
L(z) \eql -\txt{\frac{1}{2}} \int_0^z \ dy \left( 
\frac{\log y}{1-y}+ \frac{\log(1-y)}{y} \right)  \,.
\end{equation}
The quantities $\xi_i$ which appear in Eq.\ \eqref{sec3eq11ea} are
solutions to the central charge equations
\begin{equation} \label{sec3eq13ea}
\xi_i \eql \prod_j (1-\xi_j)^{\bK_{ij}} \,.
\end{equation}
For the abelian quantum Hall case, we have two matrices
$\bK_\phi$ and $\bK_e$ and we need the solutions 
$\xi_i$ and $\eta_i$ of the equations 
\begin{equation} \label{sec3eq5ea}
\xi_i \eql \prod_{j=1}^n\  (1-\xi_j)^{(\bK_\phi)_{ij}} \,, \qquad
\eta_i \eql \prod_{j=1}^n\  (1-\eta_j)^{(\bK_e)_{ij}} \ .
\end{equation}
By virtue of the duality, these solutions are related 
by a simple equation: $\eta_i = 1-\xi_i$. This leads to
\begin{equation} \label{sec3eq14ea}
\sum_i L(\xi_i) + \sum_i L(\eta_i) \eql \sum_i \left( L(\xi_i) + L(1-\xi_i) 
\right) \eql n L(1) \eql  n \, \frac{\pi^2}{6} \,.
\end{equation}
So in the abelian case, we correctly find that the central charge is just 
given by the number of species in the theory, $c_{\text{CFT}} = n$. 

\section{Non-abelian exclusion statistics}

In this section, we focus on $K$-matrices and statistics
matrices for non-abelian quantum Hall states. We shall first
introduce new types of particles, pseudoparticles and
composite particles, and explain the role
they play in the non-abelian case. We also extend the
notion of duality to the non-abelian case. After that
we discuss various aspects (filling factors and shift
map) of the quantum Hall data $\bK$, $\bt$, $\bs$ and
$\bj$ in the non-abelian case.

Among the new particles that appear in non-abelian theories
are so called `composite' particles in the electron sector. 
These will show up as particles which have multiple electron 
charges. We introduce an integer label $l_i$ for an
order-$l_i$ composite particle of charge $(\bt_e)_i=-l_i$.

In the quasihole sector, we encounter so called 
pseudoparticles, which do not carry any energy, but rather 
act as a book-keeping device that keep track of
`internal degrees of freedom' of the physical quasiholes.
Pseudoparticles were first introduced in the TBA analysis
of integrable systems with non-diagonal particle
scattering (see, e.g.\ \cite{Zam}); in the context
of exclusion statistics they have been discussed in
\cite{FK,GS,BCR,ABGS}. We assign the label $l_i=0$
to all pseudoparticles.

An important observation, first made in \cite{ABGS}, is
that the duality between the electron and quasihole sectors
naturally links the presence of composite particles in
one sector to the presence of pseudoparticles in the other.
Physically, this is a link between the pairing physics
of the non-abelian quantum Hall states and the non-abelian
statistics of their fundamental excitations.
 

\subsection{Composites, pseudoparticles and null-particles}

The presence of pseudoparticles and composite particles
calls for a slight generalization of the discussion of
the previous section. When focusing on the dependence
of the $\lambda_i$ on the energy $\epsilon$, the natural 
specialization of the generalized fugacities $z_i$ is given 
by $z_i=z^{l_i}$, with $z=e^{-\beta\epsilon}$. 
In the presence of $l_i \neq 1$, the 1-particle distribution
functions take the form [note that a composite particle 
labeled by $\epsilon$ carries energy $l_i \epsilon$]
\begin{equation} \label{sec4eq0ea}
n_i(\epsilon) \eql 
z_i \frac{\partial}{\partial z_i} \log \prod_j [\lambda_j]^{l_j} 
{}_{\big| z_i=e^{\beta(\mu_i-l_i \epsilon)}} \eql
\sum_j l_j z_j \frac{\partial}{\partial z_j} \log \lambda_i
{}_{\big| z_i=e^{\beta(\mu_i-l_i \epsilon)}} \ .
\end{equation}
With the following definition of $\latot(z)$ 
\begin{equation} \label{sec3eq6ea}
\latot(z) \eql 
\prod_i [\lambda_i(z_j=z^{l_j})]^{l_i} \ ,
\end{equation}
the central charge $c_{\text{CFT}}$ is again given by
the expression (\ref{sec3eq10ea}). We note that in the 
specialized IOW equations, with $z_i=z^{l_i}$,
the right hand side of the equations for pseudoparticles is 
equal to 1. When focusing on quantum numbers other than 
energy, such as spin, we will consider slightly more general 
versions of the quantity $\latot$.

In all examples (abelian and non-abelian) that are explicitly discussed 
in this paper, we assume a choice of particle basis such that 
$\bl_e=-\bt_e$. For the abelian quantum Hall states we further assume
that $(\bt_e)_i=-1$ for all $i$.  In the quasihole sector we specify
$(\bl_\phi)_i = \frac{1}{q_{\text{qp}}} (\bK_\phi)_{ij} (\bl_e)_j$, where
$q_{\text{qp}}$ is the smallest (elementary) charge in the quasihole
sector. [This implies that, even in the abelian case, we may treat some
of the quasiholes as composites of the most fundamental ones, thereby
generalizing the discussion of the previous section.]

Under these assumptions, we find that under duality
$\lambda_{\text{tot}}(x)$ and $\mu_{\text{tot}}(y)$ 
are related in the following way
\begin{equation} \label{sec3eq7ea}
\la_{\text{tot}}(x) \eql x^\gamma \mu_{\text{tot}}^\alpha (y)\,,
\qquad y = x^{-\beta} \,,
\end{equation}
with
\begin{equation}
\alpha = \beta = {1 \over q_{\text{qp}}}  \,,\qquad
\gamma =  {\nu \over q_{\text{qp}}^2} \,.
\end{equation}  

A clear sign of non-abelian statistics is found in the way 
the quantity $\la_i$ for physical particles depends on the fugacity 
$z_i$. Putting $z_l=1$ for all pseudoparticles,
and focusing on the small $z$ behaviour of $\lambda_i$, one 
finds
\begin{equation} \label{sec4eq1ea}
\la_i \eql 1 + \alpha_i z_i + o(z^2)\,.
\end{equation}
In the abelian case, $\alpha_i=1$, whereas in the non-abelian case 
$\alpha_i > 1$.  The factors $\alpha_i$ lead to
multiplicative factors in the Boltzmann tails of the one-particle
distribution functions for physical particles. The quantities 
$\alpha_i$ are in fact the largest eigenvalues of the fusion matrix 
\cite{BSb}, i.e., the quantum dimensions \cite{dFMS} of the conformal 
field theory associated to the quantum Hall state, and can easily 
be calculated for the cases we deal with (see Sects.\ 6 and 7.2).

In \cite{ABGS}, we presented a generalized $K$-matrix structure for
some recently proposed quantum Hall states. The proposed $K$-matrices
were identified via their role as statistics matrices for the
fundamental charged edge excitations. In the quasihole 
sector, the non-abelian statistics leads to a specific set
of pseudoparticles and an associated statistics matrix 
$\bK_\phi$ \cite{GS,BCR}. The  matrix $\bK_e$, related to 
$\bK_\phi$ by the duality $\bK_e=\bK_\phi^{-1}$, refers to particles 
which are identified as composites of the fundamental electron-like 
excitation. {}From the point of view of the 
wave functions for the non-abelian quantum 
Hall states \cite{MR,RR,AS}, the presence of composite excitations
is very natural. This is because the non-abelian states show a 
behaviour which is called {\em clustering} (of order $k$, where $k$ is 
a label of the states \cite{RR,AS}). This order-$k$ clustering means
that up to $k$ particles can come to the same position, without
making the wave function zero, whereas, as soon as $k+1$
particles are located at the same positions, the wave function
becomes identically zero. In \cite{GWW,ABGS} it was argued that the wave
functions which show pairing (at $k=2$), are related (in the 
non-magnetic limit, i.e.\ in the limit of $\nu \rightarrow \infty$)
to BCS superconductivity.

Composite particles are identified as particles whose generalized
fugacities are specific combinations of the generalized
fugacities of other particles, i.e., all quantum numbers of
composite particles are completely determined in terms of the quantum
numbers of their constituents.  It has been shown in \cite{BCR} that 
particular kinds of composite particles, so-called null-particles, 
accounting for the null-states in the quasiparticle Fock spaces,
are often needed to interpret the system in terms of Haldane's 
exclusion statistics or, equivalently, to write the partition function
in UCPF form (see also Sect.\ 5.2). 

We now turn to the computation of the central charge $c_{\text{CFT}}$
the non-abelian case. It was shown in \cite{BCR}, that the presence of
pseudoparticles leads to a simple correction term that is
subtracted from the abelian result $c_{\text{CFT}}=n$. 
For the pseudoparticles, 
a system of equations like Eq.\ \eqref{sec3eq5ea} can be written down
\begin{equation} \label{sec4eq2ea}
\xi'_i \eql  \sideset{}{'}\prod_j (1-\xi'_j)^{\bK_{ij}} \,,
\end{equation}
where the prime indicates that the product is restricted to 
pseudoparticles. The correction term is given by a 
sum over the dilogarithm of the solutions of \eqref{sec4eq2ea},
leading to 
\begin{equation} \label{sec4eq3ea}
c_{\text{CFT}} \eql n - {\frac{6}{\pi^2}} 
  \sideset{}{'}\sum_i\ L(\xi'_i) \,.
\end{equation}

\subsection{On filling factors}

Up to now, we merely asserted that the statistics matrices 
$\bK$ can also serve as (generalized) $K$-matrices for
non-abelian quantum Hall states. To make this statement more 
clear, we will now investigate how some of the `$K$-matrix
results' for abelian quantum Hall states generalize to the
non-abelian case. In this derivation, we make the assumption 
that the pseudoparticles do not carry charge or spin. In all 
cases that are explicitly considered in Sects.\ 6 and 7 this 
assumption holds in the simplest formulation. If pseudoparticles 
do carry spin or charge, the formulas we obtain below need to 
be modified.

Let us start with the filling factor corresponding to state
which is described by the IOW-equations, for a statistics matrix 
$\bK_e$, charge vector $\bt_e$, and labels $\bl_e=-\bt_e$.  
We couple the system to an electric field by taking 
$y_i = y^{-(\bt_e)_i}$. [This is when the orientation of the
electric field is such that the response is carried by the negatively
charged excitations.] The large $y$ (i.e.\ low temperature) behaviour
of the IOW-equations \eqref{sec3eq1ea}
is then given by the following set of relations
\begin{equation} \label{sec4eq4ea}
\prod_j \mu_j^{(\bK_e)_{ij}} \sim y^{-(\bt_e)_i} \ ,
\end{equation}
which imply, when $\bK$ is symmetric (which is assumed throughout 
the paper)
and invertible
\begin{equation} \label{sec4eq5ea}
\mutot \eql \prod_i \mu_i^{-(\bt_e)_i} \sim
y^{\bt_e \cdot \bK_e^{-1} \cdot \bt_e} \,.
\end{equation}
Because the left hand side of Eq.\ \eqref{sec4eq5ea} in the 
$T\to0$ limit determines the filling factor $\nu$ through
$\mutot \sim y^{\nu}$,
we find the well-known formula 
\begin{equation}  \label{sec4eq6ea}
\nu \eql \bt_e \cdot \bK_e^{-1} \cdot \bt_e \ .
\end{equation}
For the opposite orientation of the electric field,  a similar 
expression is obtained by starting from the 
$K$-matrix for the (positively charged) quasiholes
\begin{equation} \label{sec4eq7ea}
\nu \eql \bt_\phi \cdot \bK_\phi^{-1} \cdot \bt_\phi \,.
\end{equation}
This result could also have been obtained by using 
Eq.\ \eqref{sec4eq6ea} and the transformation properties of 
$\bK_e$ and $\bt_e$ under duality. We remark that the 
above derivations explicitly assume that only the physical 
particles respond to the electric field, i.e., that all pseudoparticles 
are neutral.

Let us now turn to the spin Hall conductance, and the
corresponding spin filling factor. 
The derivation of the corresponding spin filling 
factor
\begin{equation} \label{spinfill}
\sigma \eql \bs_e \cdot \bK_e^{-1} \cdot \bs_e \,,
\end{equation}
goes along the same lines as the derivation of the electron filling
factor. As an extra step, one needs to relate the fugacities
of the spin up and down particles by
$y_\ua = 1/y_\da = z$. This results in
\begin{equation} 
\prod_i \mu_i^{(\bs_e)_i} \sim
z^{\bs_e \cdot \bK_e^{-1} \cdot \bs_e} \,,
\end{equation}
leading to Eq.\ \eqref{spinfill}. It is important to note that 
this formula only holds in the cases where the pseudoparticles
in the $\phi$-sector do not carry spin. As a check on this
formula, one would like to have a procedure to obtain the 
spin filling factor directly from the wave functions, as is
possible for the electron filling factor. To do this, one
has to count the zeros of the wave function with respect to
one reference particle (of a given spin, say, up). The total 
number of zeros gives the total flux needed on the sphere
as a linear function of the total number
of electrons $N_e$. By using the relation between $N_e$ and $N_\Phi$
given in \eqref{sec2eq4ea}
one obtains the electron filling factor and the shift. 
To obtain the spin filling factor, one has to keep track
of two different types of zeros, namely those with respect 
to a particle of the same spin, and the ones with respect
to particles of the other spin. We will denote the number
of these zeros by  $N_\Phi^\ua$ and $N_\Phi^\da$
respectively. The electron and spin filling factors are obtained from
\begin{eqnarray}
N_\Phi \eql N_\Phi^\ua + N_\Phi^\da & = & \frac{1}{\nu} N_e -
\cS \,,\nonumber \\
N_\Phi^\uparrow - N_\Phi^\da & = & \frac{1}{\sigma} N_e - \cS \,.
\end{eqnarray} 
We applied this procedure to the non-abelian spin singlet 
states of \cite{AS} (the explicit form
of the wave functions will be given elsewhere \cite{ARRS}), 
and indeed found
the same results for the electron and spin filling factor
as obtained from the $K$-matrix formalism, Eq.\ \eqref{eqGAa}.
Also the electron filling factor for the Read-Rezayi states is
reproduced correctly, see Eq.\ \eqref{eqFAa}.
In addition, for both types of states we found that the shift on the 
sphere is in agreement with \eqref{sec2eq5ea} for $g=0$.

Summarizing, we have presented evidence that duality relations
\begin{equation} \label{eqDa}
\bK_\phi \eql \bK_e^{-1} \,,\qquad
\bt_\phi \eql - \bK_e^{-1} \cdot \bt_e\,, \qquad
\bs_\phi \eql - \bK_e^{-1} \cdot \bs_e\,, \qquad
\bj_\phi \eql - \bK_e^{-1} \cdot \bj_e\,.
\end{equation}
are applicable to both abelian and non-abelian quantum Hall states, 
and that the expressions (\ref{sec2eq2ea}) for the filling factors 
$\nu$ and $\sigma$ apply to the non-abelian case, in a
formulation where pseudoparticles do not carry spin or charge.

\subsection{Shift map}

Suppose we have a fractional quantum Hall system described by the data 
$(\bK_e,\bt_e,\bs_e,\bj_e)$. We can then construct a family of fractional
quantum Hall systems, parametrized by $M\in\ZZ_+$, by applying the 
`shift map' $\cS_M$ introduced in \cite{FKST}. In the cases we 
consider, $M$ odd (even) corresponds to a fermionic (bosonic) state
respectively. At the level of
trial wave functions $\Psi(z)$, $\cS_M$ simply acts as
a multiplicative Laughlin factor $\prod_{i<j} (z_i-z_j)^M$.  
Thus, $\cS_M$ increases the number of flux quanta by
\begin{equation}
N_\Phi \mapsto N_\Phi + M(N_e-1) \eql ( \frac{1}{\nu} +M)N_e - (\cS+M) \,,
\end{equation}
i.e.,
\begin{equation} \label{eqDCe}
\nu^{-1} \mapsto \nu^{-1} + M \,,\qquad \sigma \mapsto \sigma \,, \qquad
\cS \mapsto \cS +M \,.
\end{equation}
In fact, $\cS_M$ acts on the fqH data $(\bK_e,\bt_e,\bs_e,\bj_e)$ as 
\begin{eqnarray} \label{eqDCf}
\cS_M \bK_e & \eql & \bK_e + M \bt_e \bt_e \,,\nonumber \\
\cS_M \bt_e & \eql & \bt_e \,,\nonumber \\
\cS_M \bs_e & \eql & \bs_e\,,\nonumber \\
\cS_M \bj_e & \eql & \bj_e + \txt{\frac{M}{2}} \bt_e \,.
\end{eqnarray}
One easily checks that \eqref{eqDCf}, together with \eqref{sec4eq6ea}, 
leads to the shift in $\nu^{-1}$ as given in \eqref{eqDCe}.  
By duality \eqref{eqDa} one obtains
\begin{eqnarray} \label{eqDCg}
\cS_M \bK_\phi & \eql & \bK_\phi  - \txt{\frac{M}{1+\nu M}} \bt_\phi
    \bt_\phi \,,\nonumber \\
\cS_M \bt_\phi & \eql & \txt{\frac{1}{1+\nu M}} \bt_\phi \,,\nonumber \\
\cS_M \bs_\phi & \eql & \bs_\phi \,,\nonumber \\
\cS_M \bj_\phi & \eql & \bj_\phi - 
\txt{\frac{M}{2}\left(\frac{\nu \cS -1}{1+\nu M} \right)} \bt_\phi \,.
\end{eqnarray}
A few remarks should be made. By using the duality
\eqref{eqDa}, one actually finds for the action of the shift map
on $\bs_\phi$: $\cS_M \bs_\phi = \bs_\phi +
\frac{M(\bt_\phi \cdot \bs_e)}{1+\nu M} \bt_\phi$. However, the shift
map is only supposed to act on the charge component of the particles,
thus we would like to demand that $\cS_M \bs_\phi = \bs_\phi$.
Therefore, for consistency, we {\it require} 
\begin{equation} \label{eqDCh}
\bt_\phi \cdot \bs_e \eql -\bt_e \cdot \bK_e^{-1} \cdot \bs_e \eql 0 \,,
\end{equation}
leading to \eqref{eqDCg}. Of course, relation \eqref{eqDCh} is just
the statement that for spin singlet states there should be a 
$\ZZ_2$ symmetry $(\bt_e,\bs_e) \mapsto (\bt_e,-\bs_e)$.  
Eq.\ \eqref{eqDCh}
is fulfilled for all our examples (if we take $\bs_e = 0$
for the spin polarized states). 
Although, in general, $\bj_e$ has to be treated as an independent 
variable, for the examples discussed in Sects.\ 6 and 7 all formulas
are consistent with the relation $\bj_e = \bs_e + (\cS / 2(1-g))\bt_e$.

In this paper we will be mainly concerned with fractional quantum Hall
systems corresponding to conformal field theories $\wh{\mathfrak
g}_{k,M}$ which are deformations of the conformal field theory based
on the affine Lie algebra $\wh{\mathfrak g}_k$ at level $k$.  The
$\wh{\mathfrak g}$-symmetry greatly simplifies the determination of
the fqH data $(\bK_e,\bt_e,\bs_e,\bj_e)$ for $\wh{\mathfrak{g}}_k$.
The fqH data for $(\wh{\mathfrak g})_{k,M}$ are then simply obtained
by applying the shift operator $\cS_M$ as in \eqref{eqDCf}.
The action of the shift map can be visualized as follows.  Charge is 
usually identified with a particular direction in the weight lattice 
of $\mathfrak g$.  The degrees of freedom associated to this direction 
can be represented by a chiral boson compactified on a circle of 
some radius $R$.  The shift map $\cS_M$ has the effect of rescaling
the radius $R$ while keeping all other directions in the weight diagram
fixed.

\subsection{Composites}

The description of a physical system in terms of a
set of $n$ quasiparticles with mutual exclusion statistics
given by a matrix $(\bK_{ij})_{1\leq i,j\leq n}$ is not unique.  
In particular one may extend the number of quasiparticles by
introducing composites as we will now explain.

Consider the IOW-equations \eqref{sec3eq1ea} with 
\begin{equation}\label{eqDDa}
\bK \eql 
\begin{pmatrix}  a_{11} & \ldots & a_{1n}  \\
\vdots &  & \vdots   \\
a_{n1}  & \ldots & a_{nn} \end{pmatrix} \,, \qquad \qquad
\bz \eql \begin{pmatrix} z_1 \\ \vdots \\ z_n \end{pmatrix} \,.
\end{equation}
If we define the operation $\cC_{ij}$, corresponding to adding
a composite of the quasiparticles $i$ and $j$ to the system,
by 
\begin{equation} \label{eqDDb}
\cC_{ij}\bK \eql 
\begin{pmatrix}  a_{11} & \ldots &  & a_{1n} & \vdots & 
a_{1i}+a_{1j} \\
\vdots &  & & \vdots & \vdots & \vdots \\
      &&a_{ij}+1 && \vdots & \\
      &&      && \vdots & \\
  & a_{ji}+1 & &  & \vdots & \\
      &&      && \vdots & \\
 a_{n1} & \ldots & & a_{nn} & \vdots & a_{ni}+a_{nj}\\
\hdotsfor[1.5]{4} &\vdots & \hdotsfor[1.5]{1}  \\
a_{i1}+a_{j1} & \ldots &      & a_{in}+a_{jn} & 
\vdots & a_{ii} + 2 a_{ij} + a_{jj}  \end{pmatrix} \,,
\end{equation}
and 
\begin{equation}\label{eqDDc}
\cC_{ij} \bz \eql (\bz_1, \ldots, \bz_n; \bz_i\bz_j) \,,
\end{equation}
such that, in particular, 
\begin{eqnarray}\label{eqDDd}
\cC_{ij} \bt & \eql & (\bt_1, \ldots, \bt_n; \bt_i+\bt_j) \,,\nonumber\\
\cC_{ij} \bs & \eql & (\bs_1, \ldots, \bs_n; \bs_i+\bs_j) \,,
\end{eqnarray}
then the two systems are equivalent, at least at the level of 
thermodynamics.  The solutions $\{\la_i\}$ to the IOW-equations 
defined by $(\bK,\bz)$  and $\{\la_i'\}$ defined by $(\bK',\bz') =
(\cC_{ij}\bK,\cC_{ij}\bz)$ are simply related by 
\begin{align}
\la_i' & \eql \frac{\la_i+\la_j-1}{\la_j} \,, & 
\la_j' & \eql \frac{\la_i+\la_j-1}{\la_i} \,,\nonumber\\
\la_{n+1}' & \eql \frac{\la_i\la_j}{\la_i+\la_j-1} \,, &
\la_k' & \eql \la_k\,,\qquad(k\neq i,j,n+1) \,.  
\end{align}
Note that, in particular, it follows $\la_i=\la_i'\la_{n+1}'$ and
$\la_j=\la_j'\la_{n+1}'$ such that $\latot = \latot'$.
Also, from $\la_i=\la_i'\la_{n+1}'$ and $\la_j=\la_j'\la_{n+1}'$ one
sees that the original one-particle partition functions for $i$ and
$j$, receive contributions from the new particles
$i$ and $j$, respectively, as well as from the composite 
particle $n+1$. The operation $\cC_{ij}$ has the effect that 
states in the spectrum containing both particles $i$ and $j$ get less
dense (their mutual exclusion statistics is bumped up by $1$), while
the resulting `gaps' are now filled by the new composite particle.

A consistency check on the equivalence of the systems described by 
$(\bK,\bz)$ and $(\bK',\bz')$ is the fact that both lead
to the same central charge as a consequence of the five-term
identity for Rogers' dilogarithm (see \cite{BCR}).

Finally, note that the shift map $\cS_M$ of Eq.\ \eqref{eqDCf} 
and composite operation
$\cC_{ij}$ of Eqs.\ \eqref{eqDDb} and \eqref{eqDDd} commute, i.e.\ 
\begin{equation}
\cS_M \, \cC_{ij} \eql \cC_{ij} \,\cS_M \,,
\end{equation}
as one would expect.

\section{The UCPF and exclusion statistics}

\subsection{Quasiparticle basis and truncated partition function}

Quasiparticles in two dimensional conformal field theories are
represented by so-called chiral vertex operators $\phi^{(i)}(z)$ 
that intertwine between the irreducible representations 
of the chiral algebra.  Given a set of quasiparticles $\phi^{(i)}(z)$,
$i=1,\ldots,n$, one has to determine a basis for the Fock space created 
by the modes $\phi^{(i)}_{-s}$, i.e., a maximal, linearly independent 
set of vectors
\begin{equation} \label{eqEAa}
\phi^{(i_N)}_{-s_N} \ldots \phi^{(i_2)}_{-s_2}\phi^{(i_1)}_{-s_1}
|\omega\rangle \,,
\end{equation}
with suitable restrictions on the mode sequences $(s_1,\ldots,s_N)$
(which may depend on the `fusion paths' $(i_1,\ldots,i_N)$),
as well as a set of vacua $|\omega\rangle$ (see \cite{BSb,BCR}
for more details).  The partition function $Z(\bz;q)$ is then defined 
by 
\begin{equation} \label{eqEAab}
Z(\bz;q) \eql \tr\left( (\prod_i z_i^{N_i}) q^{L_0} \right)\,,
\end{equation}
where the trace is taken over the basis \eqref{eqEAa} and
$N_i$ denotes the number operator for quasiparticles of type $i$
while $L_0=\sum_i s_i$ for a state of type \eqref{eqEAa}.
During this discussion on the UCPF, we use the following, 
in the literature standard notation
$q=e^{-\beta \epsilon_0}$, where $\epsilon_0$ is 
some fixed energy scale, and $z_i = e^{\beta \mu_i}$.

Exclusion statistics in conformal field theory can be studied by
means of recursion relations for truncated partition functions
\cite{Sca}.  Truncated partition functions $P_\bL(\bz;q)$, for
$\bL=(L_1,\ldots,L_n)$, are defined by taking the partition function 
of those states \eqref{eqEAa} where all the modes $s$ for quasiparticles 
of species $i$ satisfy $s\leq L_i$.  By definition, for large $\bL$,
we will have (see \cite{BSb,BCR} for more details)
\begin{equation} \label{eqEAb}
P_{\bL+\bfe_i}(z;q)/P_{\bL}(z;q) \sim \lambda_i(z_iq^{L_i}) \,,
\end{equation}
where $\bfe_i$ denotes the unit vector in the $i$-direction.
In particular, if the generalized fugacities $z_i$ are given by 
$z_i =z^{l_i}$, for some fixed $z$, and the 
quasiparticle modes are truncated by $L_i=l_i L$, then we find,
using \eqref{sec3eq6ea}
\begin{equation} \label{eqEAc}
P_{L+1}(z;q)/P_L(z;q) \sim \latot(zq^L) \,,
\end{equation}
where $P_L(z;q)= P_{l_1L,l_2L,\ldots,l_nL}(z_i=z^{l_i};q)$.
Thus, given a set of recursion relations for the truncated partition 
functions $P_{\bL}(z;q)$, one derives algebraic equations for the 
one-particle partition functions $\lambda_i(z)$ by 
taking the large $\bL$ limit.  In particular one can find an equation
for $\latot(z)$ from $P_L(z;q)$ by using \eqref{eqEAc}. 
For all conformal field theories that have been studied this way it turns 
out that one finds agreement between these $\lambda$-equations
and the IOW-equations \eqref{sec3eq1ea} corresponding to a specific 
statistics matrix $\bK$ (see, in particular, \cite{BSb}).

\subsection{The universal chiral partition function}

Based on many examples, it has become clear that the characters of the 
representations of all conformal 
field theories can be written in the form of, what is now known as,
a universal chiral partition function (UCPF) (see in particular,
Ref.\ \cite{BM} and references therein)
\begin{equation} \label{eqEBaa}
Z(\bK;\bQ,\bu|\bz;q) \eql
\sideset{}{'}\sum_{\bm} \left( \prod_i z_i^{m_i} \right)
q^{ \frac{1}{2} \bm \cdot \bK \cdot \bm + \bQ\cdot \bm}
\prod_i \qbin{((\id-\bK)\cdot\bm + \bu )_i}{m_i} \,,
\end{equation}
where $\bK$ is a (rational) $n\times n$ matrix, $\bQ$ and $\bu$ are
certain $n$-vectors and the sum over $m_1,\ldots,m_n$, is over the  
nonnegative integers subject to some restrictions (which, throughout 
this paper, are taken to be
such that the coefficients in the $q$-binomials are integer).
The $q$-binomial (Gaussian polynomial) is defined by 
\begin{equation} \label{eqEBab}
\qbin{M}{m} \eql \frac{\qn{M}}{\qn{m}\qn{M-m}}\,,\qquad
\qn{m} \eql \prod_{k=1}^m (1-q^k) \,.
\end{equation}
The vectors $\bQ$ and $\bu$ as well as the restrictions on the summation 
variables, will in general depend on the particular representation 
of the conformal field theory, while $\bK$ is independent 
of the representation.
To write the conformal characters in the form \eqref{eqEBaa} may
require introducing null-quasiparticles which account
for null-states in the quasiparticle Fock space \cite{BCR}.
The null-quasiparticles are certain composites, hence their fugacities
$z_i$ in \eqref{eqEBaa} are specific combinations of the fugacities of 
their constituents.

It has been conjectured that the UCPF \eqref{eqEBaa} is precisely the 
partition function \eqref{eqEAab} of a set of quasiparticles with exclusion 
statistics given by the same matrix $\bK$, where $u_i=\infty$ 
corresponds to a physical quasiparticle and $u_i<\infty$ to a pseudoparticle
\cite{GS,BCR}. This conjecture has been verified 
in numerous examples (see \cite{GS,BCR} for references).  
A convincing piece of evidence in support of this conjecture 
is the fact that the asymptotics of the character
\eqref{eqEBaa} (in the thermodynamic limit $q\to1^-$) is given by 
exactly the same formula as the one for the IOW-equations \cite{BCR}
(see also \cite{RS,DKKMM} for $z_i=1$).
In the next section we establish the correspondence in a more direct way.

For future convenience let us introduce the limiting form of the 
UCPF \eqref{eqEBa} when all $u_i\to\infty$, i.e.\ the case that all
quasiparticles are physical and the exclusion statistics is abelian
\begin{equation} \label{eqEBac}
Z_\infty(\bK;\bQ) \eql 
\sideset{}{'}\sum_{\bm} \left( \prod_i z_i^{m_i} \right)
\frac{q^{ \frac{1}{2} \bm \cdot \bK \cdot \bm + \bQ\cdot \bm}}{\prod_i
\qn{m_i}} \,.
\end{equation}
Note that the limiting UCPFs \eqref{eqEBac} are not all independent,
but satisfy (see \cite{Bo2})
\begin{equation} \label{eqEBad}
Z_\infty(\bK;\bQ) \eql Z_\infty(\bK;\bQ+\bfe_i) + z_i
q^{\frac{1}{2} \bK_{ii} + \bQ_i} Z_\infty(\bK;\bQ+\bK\cdot \bfe_i)\,,
\end{equation}
as a consequence of 
\begin{equation} \label{eqEBae}
\frac{1}{\qn{m}} \eql \frac{q^m}{\qn{m}} + \frac{1}{\qn{m-1}} \,.
\end{equation}

\subsection{Relation to exclusion statistics}

The relation between the UCPF and exclusion statistics can be 
made more explicit as follows.  Suppose the truncated partition
functions $P_\bL(\bz;q)$ are given by `finitized UCPFs' of the 
form
\begin{equation} \label{eqEBa}
P_\bL(\bz;q) \eql
\sideset{}{'}\sum_{\bm} \left( \prod_i z_i^{m_i} \right)
q^{ \frac{1}{2} \bm \cdot \bK \cdot \bm + \bQ\cdot \bm}
\prod_i \qbin{(\bL + (\id-\bK)\cdot\bm + \bu )_i}{m_i} \,,
\end{equation}
for some vectors $(\bQ,\bu)$.
Of course, the number of parameters in this expression is 
overdetermined.  Usually we think of $\bu$ as being fixed 
while the meaning of the parameters $\bL$
are determined by the cut-off scale.  We can of course absorb the $\bu$ 
by shifts in $\bL$ (in fact, in practice we often make shifts in 
the definition of $\bL$ to simplify the recursion relations).
We also remark that we have introduced 
finitization parameters $L_i$ also for the pseudoparticles 
in \eqref{eqEBa} to facilitate deriving recursion relations.
In making the identification with the truncated partition 
functions these parameters are kept at a fixed 
(usually `small' or even zero) value.

Using
\begin{equation}
\qbin{M}{m} \eql \qbin{M-1}{m} + q^{M-m} \qbin{M-1}{m-1} \,,
\end{equation}
we find that $P_\bL(\bz;q)$ satisfies the system of 
recursion relations
\begin{equation}\label{eqEBb}
P_\bL(\bz;q) \eql P_{\bL-\bfe_i}(\bz;q) + z_i q^{-\frac{1}{2} \bK_{ii}
  + \bQ_i + \bu_i + \bL_i} P_{\bL-\bK\cdot\bfe_i}(\bz;q) \,.
\end{equation}
Upon dividing by $P_\bL(\bz;q)$, setting $q=1$, taking the 
large $\bL$ limit, and using \eqref{eqEAb}, we obtain
\begin{equation} \label{eqEBzc}
1 \eql \la_i^{-1} + z_i \prod_j \la_j^{-\bK_{ji}}\,,
\end{equation}
which are equivalent to the IOW-equations \eqref{sec3eq1ea} with 
statistics matrix $\bK$.

Moreover, for any polynomial $P_\bL(\bz;q)$ satisfying the recursion relation 
\eqref{eqEBb}, the polynomial 
\begin{equation} \label{eqEBza}
Q_\bL(\bz;q) \eql \left( \prod_i z_i^{-L_i} \right)\,
q^{\frac{1}{2} \bL\cdot\bK\cdot\bL + (\bQ+\bu)\cdot\bL} 
P_{\bK\cdot \bL} (\bz;q^{-1})\,,
\end{equation} 
satisfies the recursion relations \eqref{eqEBb} with dual data 
$(\bK';\bQ',\bu',\bz')$, given by (cf.\ \eqref{sec3eq4ea})
\begin{equation} \label{eqEBzb}
\bK' \eql \bK^{-1}\,,\quad \bQ'+\bu'\eql \bK^{-1}\cdot (\bQ+\bu)\,,\quad
z_i' \eql \prod_j z_j^{-\bK^{-1}_{ij}} \,.
\end{equation} 
Thus, under the assumption that the set of finitized UCPFs \eqref{eqEBa},
for fixed $\bQ+\bu$, form a complete set of solutions to 
\eqref{eqEBb}, the dual polynomial $Q_\bL(\bz',q)$ of \eqref{eqEBza}
can again be written as a (finite) linear sum of finitized UCPFs 
with dual data \eqref{eqEBzb}.  Moreover, 
by taking the large $\bL$ limit of \eqref{eqEBza},
using Eqs.\ \eqref{eqEAb} and \eqref{eqEBzc}, 
one recovers the duality relations 
\eqref{sec3eq4ea} and \eqref{sec3eq7ea}.

The above calculation shows that, for quasiparticles whose
truncated partition function is given by an expression of the form
\eqref{eqEBa}, the thermodynamics of these quasiparticles is described by 
Haldane's exclusion statistics with statistics matrix $\bK$.
Even though many truncated characters are indeed of the form \eqref{eqEBa}
(we will encounter various examples in the remainder of this paper)
this is not the general situation.  However, in examples
it turns out that for
all recursion relations for truncated characters there is an associated
recursion relation, leading to the {\it same} $\la$-equation, which
does admit a solution of the form \eqref{eqEBa}.  The true solution to this
recursion relation will in general differ from \eqref{eqEBa}
by terms of order
$q^L$.  In a sense we can talk about the {\it universality class} 
of recursion relations as those recursion relations that give rise to 
the same $\la$-equations and hence the same exclusion statistics.

\subsection{Composites, revisited}

In Sect.\ 4.4 we have seen, at the level of 
thermodynamics (i.e.\ the IOW-equations), how to introduce composite
particles into the system in such a way that the resulting system is 
equivalent to the original system.  Due to the intimate relation 
of exclusion statistics with the UCPF, explained in Sect.\ 5.3, one
would expect that a similar construction is possible at the level of
the UCPF.  
Indeed, upon substituting the following polynomial $q$-identity
(see App.\ A for a proof)
\begin{multline} \label{eqUCbbc}
\qbin{M_1}{m_1} \qbin{M_2}{m_2} \eql \\
\sum_{m\geq0} q^{(m_1-m)(m_2-m)}
\qbin{M_1-m_2}{m_1-m} \qbin{M_2-m_1}{m_2-m} \qbin{M_1+M_2-(m_1+m_2)+m}{m} \,,
\end{multline}
into the UCPF \eqref{eqEBa} at the $(i,j)$-th entry, and subsequently 
shifting the summation variables $m_i\mapsto m_i+m$, $m_j\mapsto m_j+m$,
yields an equivalent UCPF, based on $n+1$ quasiparticles with 
data $(\cC_{ij}\bK;\cC_{ij}\bQ,\cC_{ij}\bu)$ and $\cC_{ij}\bz$, where 
\begin{eqnarray}
\cC_{ij}\bQ & \eql &  (\bQ_1, \ldots, \bQ_n; \bQ_i+\bQ_j) \,,\nonumber \\
\cC_{ij}\bu & \eql &  (\bu_1, \ldots, \bu_n; \bu_i+\bu_j) \,,
\end{eqnarray}
while $\cC_{ij}\bK$ and  $\cC_{ij}\bz$ are defined in Eqs.\ \eqref{eqDDb}
and \eqref{eqDDc}, respectively.
Various limiting forms of \eqref{eqUCbbc}, relevant to introducing a
composite of two physical particles or one physical particle and 
one pseudoparticle, are given in App.\ A as well.

\section{$\mathfrak{sl}_2$: $K$-matrices for non-abelian spin polarized
states}

In this section we discuss a family of non-abelian spin polarized
fractional quantum Hall systems with underlying conformal field 
theory $(\wh{\mathfrak{sl}_2})_{k,M}$ and filling factor 
\begin{equation} \label{eqFAa}
\nu_{k,M} \eql \frac{k}{kM+2} \,.
\end{equation}
For $k=2$ these systems, the so-called $q$-Pfaffians
(where now $q=1/\nu=M+1$),
were introduced in \cite{MR} while the generalizations to $k>2$ 
were introduced in \cite{RR}.  
The system contains a single quasihole $\phi$, with
charge $1/(kM+2)$ and an electron operator 
$\Psi$ with charge $-1$.  At the $(\wh{\mathfrak{sl}_2})_k$-point
(i.e.\ $M=0$) the quasihole operator $\phi$ has $\mathfrak{sl}_2$-weight
$\alpha/2$, where $\alpha$ is the (positive) root of $\mathfrak{sl}_2$
and corresponds to one component of the chiral vertex operator
transforming in the spin-$1/2$ representation 
(`spinon', see \cite{Hala,BPS,BLSa,BLSb}),
while the electron operator has weight $-\alpha$ and corresponds to the
current $J_{-\alpha}$.  For general $M$ the charge lattice has to be 
stretched.

The fqH data $(\bK_e,\bt_e)$ and their duals $(\bK_\phi,\bt_\phi)$ 
for $k=1$ (corresponding to the abelian spin polarized
Laughlin states with $\nu = 1/(M+2)$ \cite{Laug})
were discussed in \cite{ES} and for $k=2$ 
(the $q$-Pfaffian) in \cite{ABGS}.  Here we discuss the generalization 
(see also \cite{GS}) to arbitrary $k$, corresponding to the Read-Rezayi
states \cite{RR}.

As indicated before, we analyze the conformal field theory 
$(\wh{\mathfrak{sl}_2})_{k,M}$ by first analyzing the affine 
Lie algebra point $M=0$ and subsequently applying the shift map 
to obtain the result for general $M$.

The exclusion statistics and UCPF for the doublet of 
spinon operators in $(\wh{\mathfrak{sl}_2})_{k}$ were studied in
\cite{BLSb,FrS,GS,BCR}. It turns out that in this case we need 
$k-1$ additional charge- and spin neutral pseudoparticles.  Omitting 
the negative isospin spinon, we find (see, in particular, \cite{GS,BCR})
\begin{equation} \label{eqFAaa}
\bK_\phi  \eql \begin{pmatrix}
1 & -\txt{\frac{1}{2}} & &&&   \vdots &   \\
 -\txt{\frac{1}{2}} & 1 & -\txt{\frac{1}{2}} &  &   & \vdots & \\
 & \ddots & \ddots & \ddots &  & \vdots & \\
 & & -\txt{\frac{1}{2}} & 1 & -\txt{\frac{1}{2}} &  \vdots & \\
  &  &        & -\txt{\frac{1}{2}}  & 1  & \vdots &  -\txt{\frac{1}{2}} \\
\hdotsfor[1.5]{7} \\
 &  &        &    &  -\txt{\frac{1}{2}}  & \vdots & \txt{\frac{1}{2}}
\end{pmatrix}\,, \qquad
\bt_\phi \eql \begin{pmatrix} 0 \\  \vdots \\ 0 \\ 
\txt{\frac{1}{2}} \end{pmatrix} \,,
\end{equation}
leading, with \eqref{sec4eq7ea}, to a filling factor of $\nu=k/2$ in 
accordance with \eqref{eqFAa}.

The data for arbitrary $M$ now follow by applying the shift map 
$\cS_M$ of \eqref{eqDCg}, i.e.\ 
\begin{equation} \label{eqFAb}
\bK_\phi^M  \eql \cS_M \bK_\phi \eql \begin{pmatrix} 
  & & & \vdots &   \\
  & \txt{\frac{1}{2}} \bA_{k-1} & & \vdots & \\
  & & & \vdots & -\txt{\frac{1}{2}} \\
 \hdotsfor[1.5]{5} \\
  & & -\txt{\frac{1}{2}} & \vdots & \txt{\frac{(k-1)M+2}{2(kM+2)}}
\end{pmatrix} \,,\qquad
\bt_\phi^M \eql \begin{pmatrix} 0 \\  \vdots \\ 0 \\ 
\txt{\frac{1}{kM+2}} \end{pmatrix} \,,
\end{equation}
where, in order to simplify the notation, we have introduced the 
Cartan matrix $\bA_{k-1}$ of $\mathfrak{sl}_k$ (cf.\ \eqref{eqAPBc}).
One verifies that \eqref{sec4eq7ea} is satisfied.
The IOW-equations, determining the exclusion statistics of the 
quasiholes, can now be explicitly written down.  E.g., for the
$q$-Pfaffian ($k=2$)  the following equation for $\latot$ 
easily follows from \eqref{sec3eq1ea}, in agreement with \cite{Scb}
\begin{equation} \label{eqFAba}
(\latot-1)(\latot^{\frac{1}{2}}-1) 
  \eql x^2 \latot^{\frac{3M+2}{2(M+1)}} \,.
\end{equation}

The small $x$ behaviour of $\latot$ for general $k$ was
obtained from the IOW-equations in \cite{BSb}, with the result
\begin{equation} \label{eqFAg}
\latot (x) \eql 1 + \alpha_k x + o(x^2) \,, \qquad
\alpha_k \eql 2 \cos \left( \frac{\pi}{k+2} \right) \,. 
\end{equation} 
It was argued that the factors $\alpha$ can also be obtained 
as quantum dimension of the appropriate CFT. It is easily 
checked that the small $x$ behaviour of $\latot$ in \eqref{eqFAba}
indeed satisfies \eqref{eqFAg} for $k=2$. Similar equations
for $\latot$ with $k=3,4$ were given in \cite{BSb}.

To determine the fqH data $(\bK_e,\bt_e)$ in the electron sector 
we observe that the electron operator $\Psi(z)$ is identified with 
$J_{-\alpha}(z)$.  By acting with the negative modes of $J_{-\alpha}(z)$
on the lowest weight vector in the lowest energy sector of some 
integrable highest weight module $L(\Lambda)$ 
at level $k$, one obtains what is 
known as the principal subspace $W(\Lambda)$ of $L(\Lambda)$ 
(or, rather, the reflected principal subspace).
It is known that the character of the principal subspace can be written
in the UCPF form \cite{FS,Geo} (see App.\ B for a brief summary of the 
results for $(\wh{\mathfrak{sl}_n})_k$).  For $(\wh{\mathfrak{sl}_2})_k$
this requires, besides the electron operator $\Psi$ itself, clusters
of up to $k$ electron operators.  The corresponding $K$-matrix is given by
the $k\times k$ matrix $\bK_e = 2\, \bB_k$ where $(\bB_k)_{ij}= \min(i,j)$
(see \eqref{eqAPBd}),
while $\bt_e=-(1,2,\ldots,k)$.  Applying the shift map \eqref{eqDCf} 
thus gives
\begin{equation} \label{eqFAc}
\bK_e^M \eql \begin{pmatrix}
M+2    & 2M+2     & \ldots & kM+2    \\
2M+2   & 2(2M+2)  & \ldots & 2(kM+2) \\
\vdots & \vdots   & \ddots & \vdots  \\
kM+2   & 2(kM+2)  & \ldots & k(kM+2) 
\end{pmatrix} \,,\qquad
\bt_e^M \eql -\begin{pmatrix} 1 \\ 2 \\ \vdots \\ k \end{pmatrix} \,.
\end{equation}
One easily verifies that the data $(\bK_\phi,\bt_\phi)$ and 
$(\bK_e,\bt_e)$ are indeed related by the duality relations \eqref{eqDa},
and that Eqs.\ \eqref{sec4eq6ea} and \eqref{sec4eq7ea} are satisfied.

Moreover, the resulting IOW-equations for $\mutot=\mu_1\mu_2^2$ in
case of the $q$-Pfaffian are given by 
\begin{equation} \label{eqFAd}
(\mutot^{2(M+1)} - y^2) (\mutot^{M+1} - y ) \eql \mutot^{3M+2} \,,
\end{equation}
which are indeed related to \eqref{eqFAba} by the duality relations 
\eqref{sec3eq7ea}.
Explicitly, 
\begin{equation} \label{eqFAe}
\latot(x) \eql y^{-2} \mutot^{2(M+1)}(y) \,,\qquad y\eql x^{-2(M+1)} \,.
\end{equation}

Finally, in order to show that the quasihole-electron system
based on $\bK=\bK_\phi^M\oplus\bK_e^M$ gives a complete description of 
the $(\wh{\mathfrak{sl}_2})_{k,M}$ conformal field theory, we have to
show that the chiral character of the latter can be written in terms
of a (finite) combination of UCPF characters based on 
$\bK_\phi^M\oplus\bK_e^M$.  This is indeed possible and discussed in 
App.\ C.  Here we suffice to remark that the central charge, 
related to the asymptotic behaviour of the characters, works out 
correctly.  Indeed, using standard dilogarithm identities one 
finds with \eqref{sec4eq3ea}
\begin{equation} \label{eqFAf}
c_\phi + c_e \eql \frac{3k}{k+2} \,,
\end{equation}
which equals the central charge of $(\wh{\mathfrak{sl}_2})_{k,M}$.

The above description of the Read-Rezayi states has an interesting 
application, namely the identification of a particle which acts
as a supercurrent in the non-magnetic limit.  This identification
was made in \cite{ABGS},
to which we refer for a more detailed discussion.  
We use the variable 
$q=1/\nu=M+k/2$, in terms of which the non-magnetic limit corresponds 
to $q \to 0$. In this limit, all the statistics parameters of
the largest composite (with charge $-k$), go to zero, while the 
statistics parameters of the quasihole diverge. This is 
easily seen when one writes the statistic matrices \eqref{eqFAc} 
and \eqref{eqFAb} in terms of $q$. For these quantum Hall 
states the fundamental flux quantum is $h/ke$,
because of the order-$k$ clustering.
Upon piercing a quantum Hall state with this
amount of flux, a quasihole with charge $e/kq$ is 
formed. This follows from the fact that the filling factor is
$e^2/qh$ in physical units.
For $q \geq 1/k$ this is the lowest charge possible
and the electron like excitations correspond to multiple
insertions of the flux quantum.
This situation changes when we take the limit $q \to 0$. Following
\cite{ABGS}, we take $q=1/N$, with $N$ a large integer. 
The largest composite is formed by inserting an amount of flux
$-qkh/e=-kh/Ne$, thus a fraction of the 
flux quantum. The maximal occupation with this particle (in
absence of other particles) is
$n_{\text{max}} = 1/k^2 q = N/k^2$.
Thus the maximal amount of flux that can be screened by this
type of composites is $(-kh/Ne) (N/k^2) = -h/ke$,
which is precisely the flux quantum. In conclusion we find that
in the non-magnetic limit, the largest composite has bosonic
statistics, and can screen an amount of flux up to the 
flux quantum. This clearly resembles the behaviour of the 
supercurrent  in BCS superconductors.

\section{$\mathfrak{sl}_3$: $K$-matrices for non-abelian spin singlet
states}

In \cite{AS} a family of non-abelian spin singlet (NASS)
states $\Psi_{k,M}$ trial wave functions with filling factors
\begin{equation} \label{eqGAa}
\nu_{k,M} \eql \frac{2k}{2kM+3}\,,\qquad \sigma_{k,M} \eql  2k\,,
\end{equation}
was constructed.  The system has two quasihole excitations $\{\phi_\ua,
\phi_\da\}$ with one unit of up/down spin and charge 
$1/(2kM+3)$, while the electron operators $\{\Psi_\ua,\Psi_\da\}$
have charge $-1$.
The underlying conformal field theory is $(\wh{\mathfrak{sl}_3})_{k,M}$. 
In terms of $\mathfrak{sl}_3$-weights the spin and charge assignment
in the $M=0$ case 
is as follows.  Denote the positive simple roots of $\mathfrak{sl}_3$
by $\alpha_i$, $i=1,2$ and the remaining positive non-simple root 
by $\alpha_3=\alpha_1+\alpha_2$.  Let $\epsilon_i$, $i=1,2,3$, denote the 
weights of the fundamental three dimensional irreducible representation 
$\mathbf3$ of $\mathfrak{sl}_3$ such that $\epsilon_i\cdot \epsilon_j =
\delta_{ij} - 1/3$ and $\alpha_i=\epsilon_i - \epsilon_{i+1}$, $i=1,2$, then
$\{\phi_\ua,\phi_\da\} = \{\phi^{\epsilon_1}, \phi^{\epsilon_2}\}$ while
$\{\Psi_\ua,\Psi_\da\}= \{J_{-\alpha_2},J_{-\alpha_3}\}$ (see Fig.\
\ref{figGAa}).  The charge and spin direction are identified in the 
$\mathfrak{sl}_3$ weight diagram as indicated in the figure.
For other $M$ the analogous picture is obtained by
`stretching' the charge axis.

\setlength{\unitlength}{1truecm}
\begin{figure}
\begin{picture}(10,8)(-5,-5)
\put(-5,0){\line(1,0){10}}
\put(0,-5){\line(0,1){8}}
\put(0,0){\vector(2,1){2}}
\put(0,0){\vector(-2,1){2}}
\put(0,0){\vector(1,-2){2}}
\put(0,0){\vector(-1,-2){2}}
\put(2.25,-4.5){\makebox(0,0){$-\alpha_2$}}
\put(-2.25,-4.5){\makebox(0,0){$-\alpha_3$}}
\put(2.25,1.5){\makebox(0,0){$\epsilon_1$}}
\put(-2.25,1.5){\makebox(0,0){$\epsilon_2$}}
\put(0.75,2.5){\makebox(0,0){charge}}
\put(4.5,-0.35){\makebox(0,0){spin}}
\put(-0.05,1){\line(1,0){0.1}}
\put(-0.05,-4){\line(1,0){0.1}}
\put(0.3,1){\makebox(0,0){$\frac{1}{3}$}}
\put(0.5,-4){\makebox(0,0){$-1$}}
\end{picture}
\caption{} \label{figGAa}
\end{figure}

In the following sections we analyze the fqH data for the conformal
field theory $(\wh{\mathfrak{sl}_3})_{k,M}$.  We first discuss the 
case $k=1$ (which corresponds to the abelian spin singlet Halperin
state with parameters $(M+2,M+2,M+1)$ \cite{Halp}) in some detail
and then generalize to the non-abelian case $k>1$.

\subsection{$(\wh{\mathfrak{sl}_3})_{k=1,M}$}

The exclusion statistics and UCPF character for the 
$(\wh{\mathfrak{sl}_3})_{k=1,M=0}$ conformal field theory, 
in terms of the quasiparticles $\{\phi^{\epsilon_1}, \phi^{\epsilon_2},
\phi^{\epsilon_3}\}$, were worked out in \cite{BSa,Sca,BSb,BCR}.
Specializing to the subset $\{\phi_\ua,\phi_\da\} = \{\phi^{\epsilon_1}, 
\phi^{\epsilon_2}\}$ we have 
\begin{equation} \label{eqGAb} 
\bK_\phi \eql  \txt{\frac{1}{3}} \begin{pmatrix}
2 & -1 \\ -1 & 2 \end{pmatrix} \,,\qquad
\bt_\phi \eql \begin{pmatrix}  \txt{\frac{1}{3}} \\ \txt{\frac{1}{3}}
\end{pmatrix} \,,\qquad
\bs_\phi \eql \begin{pmatrix} 1 \\ -1 \end{pmatrix} \,.
\end{equation}
With \eqref{sec4eq7ea} this leads to $\nu=2/3$ in agreement with 
\eqref{eqGAa}.
Applying the shift map \eqref{eqDCg}, the fqH data for 
$(\wh{\mathfrak{sl}_3})_{k=1,M}$ are thus given by
\begin{equation} \label{eqGAc}
\bK_\phi^M  \eql  \cS_M \bK_\phi \eql \txt{\frac{1}{2M+3}} \begin{pmatrix}
M+2 & -(M+1) \\ -(M+1) & M+2 \end{pmatrix}  \,,
\end{equation} 
while
\begin{equation} \label{eqGAe} 
\bt_\phi^M  \eql  \begin{pmatrix}  \txt{\frac{1}{2M+3}} \\ \txt{\frac{1}{2M+3}}
\end{pmatrix} \,,\qquad
\bs_\phi^M \eql \begin{pmatrix} 1 \\ -1 \end{pmatrix} \,.
\end{equation}
The IOW-equation for the total one-particle partition
function $\latot = \la_\ua \la_\da$, resulting from
\eqref{eqGAc}, is given by
\begin{equation} \label{eqGAd}
\latot - x_\ua x_\da \latot^{\frac{2M+2}{2M+3}} - (x_\ua+x_\da) 
\latot^{\frac{M+1}{2M+3}} -1 \eql 0\,.
\end{equation}

The $K$-matrix in the electron sector is determined as follows.
First of all, the principal subspace of the 
$(\wh{\mathfrak{sl}_3})_{k=1,M=0}$ integrable highest weight modules
is generated by $\{J_{-\alpha_1},J_{-\alpha_2}\}$ and has a $K$-matrix
given by (see App.\ B)
\begin{equation} \label{eqGAh}
\bK \eql \begin{pmatrix}
2 & -1 \\ -1 & 2 \end{pmatrix} \,.
\end{equation}
The electron operators $\{\Psi_\ua,\Psi_\da\}$, however, are identified with 
$\{J_{-\alpha_2},J_{-\alpha_3}\}$.  Interpreting $J_{-\alpha_3}$ as the 
composite $(J_{-\alpha_1}J_{-\alpha_2})$, we can apply the construction 
of Sect.\ 4.4 and find an equivalent $K$-matrix for the combined
$\{J_{-\alpha_1},J_{-\alpha_2},J_{-\alpha_3}\}$ system
\begin{equation} \label{eqGAi}
\bK' \eql \cC_{12} \bK \eql \begin{pmatrix}
2 & 0 & 1 \\ 0 & 2 & 1 \\ 1 & 1 & 2 \end{pmatrix} \,.
\end{equation}
Thus, we conclude that the electron fqH data are given by
\begin{equation} \label{eqGAj}
\bK_e \eql  \begin{pmatrix}
2 & 1 \\ 1 & 2 \end{pmatrix} \,,\qquad
\bt_e \eql - \begin{pmatrix} 1 \\  1 \end{pmatrix} \,, \qquad
\bs_e \eql \begin{pmatrix} 1  \\  -1 \end{pmatrix} \,.
\end{equation}
And thus, by applying the shift map 
\begin{equation} \label{eqGAl}
\bK_e^M \eql \cS_M \bK_e \eql  \begin{pmatrix}
M+2 & M+1 \\ M+1 & M+2 \end{pmatrix} \,,\qquad
\bt^M_e \eql - \begin{pmatrix}  1 \\  1 \end{pmatrix} \,.
\end{equation}
Note again that the fqH data in the electron and quasihole sectors, given
in Eqs.\ \eqref{eqGAc}, \eqref{eqGAe} and \eqref{eqGAl},
are related by the duality \eqref{eqDa}.

The IOW-equation for $\mutot=\mu_\ua\mu_\da$, resulting from 
\eqref{eqGAl}, is given by
\begin{equation} \label{eqGAm}
\mutot^{2M+3} - \mutot^{2M+2} - (y_\ua+y_\da) \mutot^{M+1} - y_\ua y_\da
\eql 0 \,,
\end{equation}
and is dual to \eqref{eqGAd} in the sense of \eqref{sec3eq7ea}.  
Explicitly,
\begin{equation} \label{eqGAn}
\latot(x_\ua,x_\da) \eql (y_\ua y_\da)^{-1} \mutot(y_\ua,y_\da)^{2M+3} \,,
\end{equation}
where
\begin{equation} \label{eqGAma}
y_\ua \eql x_\ua^{-(M+2)} x_\da^{-(M+1)} \,,\qquad
y_\da \eql x_\ua^{-(M+1)} x_\da^{-(M+2)} \,.
\end{equation}

It remains to show that the fqH data $(\bK_\phi,\bt_\phi,\bs_\phi)$
and their duals $(\bK_e,\bt_e,\bs_e)$ give a complete description of
the chiral spectrum of the $(\wh{\mathfrak{sl}_3})_{k=1,M}$ conformal 
field theory by constructing the $(\wh{\mathfrak{sl}_3})_{k=1,M}$
characters in terms of (finite) linear combinations of UCPFs based
on $\bK_e\oplus\bK_\phi$.  This is delegated to App.\ D.  Here we only
observe that, since there are no pseudoparticles, Eq.\ \eqref{sec3eq14ea}
immediately gives $c_e+c_\phi=2$ which is the correct value of the 
central charge for  $(\wh{\mathfrak{sl}_3})_{k=1,M}$.  Note also 
that $c_\phi$ and $c_e$ separately depend on $M$ and are, in general,
not simple rational numbers, e.g., for $M=0$ we have numerically
$c_e=0.6887$ and $c_\phi=1.3113$ while for $M\to\infty$ all the 
central charge
is concentrated in the $\phi$ sector.

Upon generalizing to higher levels $k>1$, it turns 
out we need an equivalent description of the system described above in 
terms of three quasihole operators, namely by adding a quasihole 
operator $\phi^{-\epsilon_3}$ of $\mathfrak{sl}_3$ weight $-\epsilon_3$,
i.e., of charge $2/3$ (for $M=0$) and spinless.  The $K$-matrix
for this system can be obtained as a submatrix of the $K$-matrix
describing quasiparticles in the $\mathbf{3}\oplus\mathbf{3}^*$ of
$\mathfrak{sl}_3$ \cite{BCR} or, equivalently, by using that
$\phi^{-\epsilon_3}$ is the composite 
$(\phi^{-\epsilon_1}\phi^{-\epsilon_2})$ \cite{BSa} and using 
\eqref{eqDDb}.  We find 
\begin{equation} \label{eqGAo}
\bK_\phi^{'M} \eql \cC_{12} \bK_\phi^M \eql \txt{\frac{1}{2M+3}}
\begin{pmatrix} 
M+2 & M+2 & 1 \\
M+2 & M+2 & 1 \\
1   &  1  & 2 
\end{pmatrix} \,,\qquad
\bt_\phi^{'M} \eql \begin{pmatrix} \txt{\frac{1}{2M+3}} \\
\txt{\frac{1}{2M+3}} \\ \txt{\frac{2}{2M+3}} \end{pmatrix} \,.
\end{equation}
In the electron sector we can similarly introduce the composite
$(J_{-\alpha_2}J_{-\alpha_3})$ and obtain
\begin{equation} \label{eqGAp}
\bK_e^{'M} \eql \cC_{12} \bK_e^M \eql 
\begin{pmatrix}
M+2 & M+2 & 2M+3 \\
M+2 & M+2 & 2M+3 \\
2M+3&2M+3 & 4M+6 
\end{pmatrix} \,,\qquad \bt'_e \eql
- \begin{pmatrix} 1 \\ 1 \\ 2 \end{pmatrix}\,.
\end{equation}
Now we observe a curiosity; while obviously the fqH data
\eqref{eqGAo} and \eqref{eqGAp} are dual, since they are equivalent to 
the dual systems given in \eqref{eqGAc} and \eqref{eqGAl}, they are not
related by the duality transformation \eqref{eqDa} because both
$\bK_\phi$ and $\bK_e$ are not invertible.  The equivalence can also
be observed at the level of the resulting IOW-equations which are 
now given by
\begin{eqnarray} \label{eqGAq}
(\latot^{\frac{1}{2M+3}} - x_\ua x_\da)(\latot - 
  x_\ua x_\da \latot^{\frac{2M+2}{2M+3}} - (x_\ua+x_\da) 
\latot^{\frac{M+1}{2M+3}} -1) & \eql&  0 \,,\nonumber \\
(\mutot^{2M+3}-y_\ua y_\da)(\mutot^{2M+3} - \mutot^{2M+2} - (y_\ua+y_\da) 
\mutot^{M+1} - y_\ua y_\da ) & \eql & 0 \,.
\end{eqnarray}
Because of the first factor the equations \eqref{eqGAq} do not
transform into eachother under \eqref{eqGAn}.  However, the physical 
solutions, which are determined by the second factor, do!
Summarizing, we conclude that it is obvious
that the notion of duality should have an extension that incorporates
non-invertible $K$-matrices.  We leave this for future investigation.

\subsection{$(\wh{\mathfrak{sl}_3})_{k,M}$}

As argued in \cite{BH,BCR}, the generalization of the results of 
the previous section to levels $k>1$ requires the addition
of $2(k-1)$ pseudoparticles incorporating the non-abelian statistics
of the quasihole operators $\{\phi_\ua,\phi_\da\}$.  Since these
pseudoparticles couple differently to $\{\phi_\ua,\phi_\da\}$ than to 
the composite particle $\phi_{\ua\da}=
(\phi_\ua\phi_\da)$ (i.e., different than 
the naive coupling given by the composite construction), it appears 
that the first construction in Sect.\ 7.1 does not generalize 
to higher levels.   

It is known that for $(\wh{\mathfrak{sl}_{n}})_{k,M=0}$
the pseudoparticles couple to the physical particles by means of the 
matrix $\bA_{n-1}^{-1}\otimes \bA_{k}$. Here
we have used the result for the restricted Kostka polynomials
as given in, e.g., \cite{BMS,DKKMM,Kir,HKKOTY} 
(see the discussion in \cite{BH} for details).
Then, by applying the shift map \eqref{eqDCg}, we obtain
\begin{equation} \label{eqGBa}
\bK^{\prime M}_\phi \eql  \begin{pmatrix}
 & & & & \vdots \\
& &  &  & \vdots &   \\
 &    & {\bA}_2^{-1}\otimes {\bA}_{k-1} &  & \vdots & 
  -\scr{\frac{2}{3}} & -\scr{\frac{2}{3}} &
  -\scr{\frac{1}{3}}   \\
  &   &  &  & \vdots & -\scr{\frac{1}{3}} &   -\scr{\frac{1}{3}} &
   -\scr{\frac{2}{3}}   \\
\hdotsfor[1.5]{8} \\
  &   & -\scr{\frac{2}{3}} & -\scr{\frac{1}{3}} & \vdots & 
\scr{\frac{(4k-1)M+6}{3(2kM+3)}} & \scr{\frac{(4k-1)M+6}{3(2kM+3)}} &
\scr{\frac{(2k-2)M+3}{3(2kM+3)}} \\
 &    & -\scr{\frac{2}{3}} & -\scr{\frac{1}{3}} & \vdots & 
\scr{\frac{(4k-1)M+6}{3(2kM+3)}} & \scr{\frac{(4k-1)M+6}{3(2kM+3)}} &
\scr{\frac{(2k-2)M+3}{3(2kM+3)}} \\
 &    & -\scr{\frac{1}{3}} & -\scr{\frac{2}{3}} & \vdots & 
\scr{\frac{(2k-2)M+3}{3(2kM+3)}} &  \scr{\frac{(2k-2)M+3}{3(2kM+3)}} &
\scr{\frac{(4k-4)M+6}{3(2kM+3)}}   \end{pmatrix} \,,
\end{equation}
where the components of $\bA_2$ refer to the quasiholes in the $\mathbf3$
and $\mathbf{3}^*$, respectively, and
\begin{equation} \label{eqGBb}
\bt_\phi \eql (\underbrace{0,0,\ldots,0}_{2(k-1)}|\txt{\frac{1}{2kM+3}},
\txt{\frac{1}{2kM+3}},\txt{\frac{2}{2kM+3}}) \,.
\end{equation}
For instance, for level $k=2$ we have
\begin{equation} \label{eqGBc} 
\bK^{\prime M}_\phi \eql \begin{pmatrix}
\frac{4}{3} & \frac{2}{3} & \vdots & -\frac{2}{3} & -\frac{2}{3} & 
-\frac{1}{3} \\
\frac{2}{3} & \frac{4}{3} & \vdots & -\frac{1}{3} & -\frac{1}{3} & 
-\frac{2}{3} \\
\hdotsfor[1.5]{6}  \\ 
-\frac{2}{3} & -\frac{1}{3} & \vdots & \frac{7M+6}{12M+9} 
& \frac{7M+6}{12M+9} & \frac{2M+3}{12M+9} \\
-\frac{2}{3} & -\frac{1}{3} & \vdots & \frac{7M+6}{12M+9} 
& \frac{7M+6}{12M+9} & \frac{2M+3}{12M+9} \\
-\frac{1}{3} & -\frac{2}{3} & \vdots & \frac{2M+3}{12M+9} 
& \frac{2M+3}{12M+9} & \frac{4M+6}{12M+9} \end{pmatrix} \,.
\end{equation}
Note that the matrix $\bK^{\prime M}_\phi$ 
of \eqref{eqGBa} is not invertible,
as was observed for $k=1$ in Sect.\ 7.1.  Thus, we cannot simply 
identify the dual sector by performing the transformation \eqref{eqDa}.

To obtain the dual sector we proceed as in Sect.\ 7.1.  We start 
with the $K$-matrix of the principal subspace spanned by $\{J_{-\alpha_1},
J_{-\alpha_2}\}$.  As discussed in App.\ B, for 
$(\wh{\mathfrak{sl}_3})_{k}$, this $K$-matrix is 
given by $\bK = \bA_2 \otimes \bB_k$ and requires, besides the currents 
$\{J_{-\alpha_1},J_{-\alpha_2}\}$ a set of $2(k-1)$ composites
\begin{equation} \label{eqGBd}
(\underbrace{J_{-\alpha_i}\ldots J_{-\alpha_i}}_{l})\,,\qquad
2\leq l\leq k\,, i=1,2\,.
\end{equation}
Starting with this matrix we introduce additional composites according to
the procedure of Sect.\ 4.4, beginning with the electron operator
$\Psi_\da=(J_{-\alpha_1}J_{-\alpha_2})$ 
(recall that $\Psi_\ua=J_{-\alpha_2}$), and continuing until all
composites 
\begin{equation} \label{eqGBdd}
(\underbrace{\Psi_\ua\ldots\Psi_\ua}_{n_\ua}
\underbrace{\Psi_\da\ldots\Psi_\da}_{n_\da})\,,\qquad
n_\ua+n_\da \leq k\,,
\end{equation}
have been introduced.  Note that the set of composites \eqref{eqGBdd},
for fixed $n_\ua+n_\da$, span a $(n_\ua+n_\da+1)$-dimensional irreducible
representation of spin $SU(2)$.  The electron $K$-matrix is then 
the $\frac{1}{2}k(k+3)\times\frac{1}{2}k(k+3)$ submatrix of the resulting 
$\bK$ obtained by omitting the composites which cannot be written
in terms of electron operators only.  Let us be illustrate this procedure
the case of $k=2$.  Starting with the principal subspace $K$-matrix
\begin{equation} \label{eqGBe}
\bK \eql \begin{pmatrix}
2 & -1 & \vdots & 2 & -1 \\
-1 & 2 & \vdots & -1 & 2 \\
\hdotsfor[1.5]{5} \\
2 & -1 & \vdots & 4 & -2 \\
-1 & 2 & \vdots & -2 & 4 \end{pmatrix} \,,
\end{equation}
we introduce, respectively, the composites 
$\Psi_\da=(J_{-\alpha_1}J_{-\alpha_2})$, 
$(J_{-\alpha_2}(J_{-\alpha_1}J_{-\alpha_1}))$,
$(J_{-\alpha_2}(J_{-\alpha_1}J_{-\alpha_2}))$, and
$(J_{-\alpha_2}((J_{-\alpha_2}(J_{-\alpha_1}J_{-\alpha_1})))$.
Then, after removing the rows and columns corresponding to
$J_{-\alpha_1}$, $(J_{-\alpha_1}J_{-\alpha_1})$ and 
$(J_{-\alpha_2}(J_{-\alpha_1}J_{-\alpha_1}))$, we obtain
\begin{equation} \label{eqGBf}
\bK_e^\prime \eql 
\begin{pmatrix}
2 & 1 & \vdots & 2 & 2 & 1 \\
1 & 2 & \vdots & 1 & 2 & 2 \\
\hdotsfor[1.5]{6} \\
2 & 1 & \vdots & 4 & 3 & 2 \\
2 & 2 & \vdots & 3 & 4 & 3 \\
1 & 2 & \vdots & 2 & 3 & 4 
\end{pmatrix} \,,\qquad 
\bt_e\eql -\begin{pmatrix}
1 \\ 1 \\2 \\2 \\2 \end{pmatrix} \,,\qquad 
\bs_e \eql \begin{pmatrix}
1 \\ -1 \\ 2 \\ 0 \\ -2 \end{pmatrix} \,.
\end{equation}
Similarly, one obtains the electron $K$-matrix for
$(\wh{\mathfrak{sl}_3})_{k,M=0}$ at higher levels, and the
generalization to arbitrary $M$ follows, as before, by applying the
shift map \eqref{eqDCf}.
Unfortunately, the procedure described above is ambiguous.  The resulting
$K$-matrix depends on the order in which the composites are taken 
as well as the precise identification of the clusters \eqref{eqGBdd} with 
the original clusters \eqref{eqGBd}, e.g., should we identify 
$(\Psi_\da\Psi_\da)$ with $(J_{-\alpha_1}(J_{-\alpha_1}(J_{-\alpha_2}
J_{-\alpha_2})))$ or $((J_{-\alpha_1}J_{-\alpha_1})(J_{-\alpha_2}
J_{-\alpha_2}))$? Ultimately, the `correct' matrix $\bK_e$ is 
selected by the requirement that the complete spectrum can be build 
out of the quasihole and electron operators or, more concretely, that
the characters of $(\wh{\mathfrak{sl}_3})_{k,M}$ can be written as a linear
combination of UCPFs based on $\bK_\phi\oplus\bK_e$. 
A nontrivial (and highly selective) check is whether the central 
charge, given by \eqref{sec4eq3ea}, works out correctly, i.e., whether 
$c_\phi + c_e = 8k/(k+3)$, for the $K$-matrices 
\eqref{eqGBa} and the `appropriate' generalization of 
\eqref{eqGBf} to higher levels 
and arbitrary $M$. We have checked this numerically for low values 
of $k$ and $M$ as well as exactly,
for all $k$, in the $M\to\infty$ limit, in which case 
the central charge is entirely concentrated in the $\phi$-sector. 
We refrain from giving the explicit matrices $\bK_e$ until we have performed
an additional simplifying reduction.

First observe that, for $k=2$, the matrix $\bK_e^\prime$ of 
Eq.\ \eqref{eqGBf}
is invertible, in contrast to the matrix $\bK^{\prime M}_\phi$ 
of \eqref{eqGBc}.
One could therefore simply have started with $\bK_e^\prime$ 
and have obtained 
the dual sector by the duality transformations (\ref{eqDa}).  
This would result in 
a $\phi$-sector, different from the one discussed above, 
with two physical quasiholes and three pseudoparticles.
Unfortunately, this procedure breaks down, in general, for higher $k$
as the matrices $\bK_e$, constructed according to the procedure outlined
above, are no longer invertible.
However, note that the matrix \eqref{eqGBf} can be reduced to an
equivalent $4\times 4$ matrix by 
inverting the composite procedure -- in this case by removing 
$(\Psi_\ua\Psi_\da)$ in the fourth column, since this column can
be created by applying $\cC_{12}$.  This procedure works for general
$k>1$ and leads to a $2k\times 2k$ electron $K$-matrix, for the composites 
\eqref{eqGBd} with either $n_\da=0$ or $n_\ua=0$ (i.e.\ we lose the 
SU(2) multiplet structure), given by 
\begin{equation}\label{eqGBg}
\bK_e \eql 
\begin{pmatrix}
2 & 0 & 2 & 0 & \cdots &  & 2 & 0 & 2 & 1 \\ 
0 & 2 & 0 & 2 & \cdots &  & 0 & 2 & 1 & 2 \\ 
2 & 0 & 4 & 0 &        &        & 4 & 1 & 4 & 2 \\ 
0 & 2 & 0 & 4 &        &        & 1 & 4 & 2 & 4 \\ 
\vdots&\vdots& &&        &        &   &   & \vdots & \vdots \\
2 & 0 & 4 & 1 &        &        & {2(k-1)}&{k-2}&{2(k-1)} & {k-1} \\ 
0 & 2 & 1 & 4 &        &        & {k-2}& {2(k-1)}&{k-1} & {2(k-1)} \\ 
2 & 1 & 4 & 2 & \cdots &  & {2(k-1)} & {k-1} & 2k & k \\ 
1 & 2 & 2 & 4 & \cdots &  &{k-1} & {2(k-1)} & k & 2k
\end{pmatrix} \,,
\end{equation}
and 
\begin{eqnarray} \label{eqGBh}
\bt_e & \eql & -(1,1;2,2;\ldots;k,k) \,,\nonumber\\
\bs_e & \eql & (1,-1;2,-2;\ldots;k,-k) \,.
\end{eqnarray}
The generalization $\bK^M_e$ 
to arbitrary $M$ follows by applying the shift
map, in this case by adding the matrix $M(\id_2 \otimes \bD)$ where 
$\id_2$ is the identity matrix in two dimensions and 
$(\bD)_{ij} = ij$ ($1\leq i,j\leq k$) (see \cite{ABGS} for an 
explicit expression in the case $k=2$).
This matrix is invertible, so we simply define 
$\bK_\phi^M=(\bK_e^M)^{-1}$.  
A convenient permutation of rows and columns of
$\bK_\phi^M$ leads to the following matrix 
\begin{equation} \label{eqGBi}
(\bK_\phi^M)^{\text{perm}} \eql
\begin{pmatrix}
& & & & & \vdots & 0 & -\frac{1}{3} \\
& & & & & \vdots & 0 & -\frac{2}{3} \\
& & {\bf A}_2^{-1} \otimes {\bf A}_{k-1} & & & \vdots & & \\
& & & & & \vdots & -\frac{2}{3} & 0 \\
& & & & & \vdots & -\frac{1}{3} & 0 \\
\hdotsfor[1.5]{8} \\
0 & 0 & & -\frac{2}{3} & -\frac{1}{3} & \vdots & 
\frac{(4k-1)M+6}{3(2kM+3)} & \frac{-M}{3(2kM+3)} \\
-\frac{1}{3} & -\frac{2}{3} & & 0 & 0 & \vdots & 
\frac{-M}{3(2kM+3)} & \frac{(4k-1)M + 6}{3(2kM+3)}  
\end{pmatrix}\,,
\end{equation}
containing two physical particles and $2(k-1)$ pseudoparticles.
Also,
\begin{eqnarray}\label{eqGBj}
\bt_\phi & \eql & (0,0;0,0;\ldots;\txt{\frac{1}{2kM+3}},
\txt{\frac{1}{2kM+3}}) 
\,,\nonumber\\
\bs_\phi & \eql & (0,0;0,0;\ldots;-1,1) \,,
\end{eqnarray}
as one would expect.
We have checked that the total central charge $c_e+c_\phi$ for 
Eqs.\ \eqref{eqGBg} and \eqref{eqGBi} works out correctly, namely
$c_e+c_\phi= 8k/(k+3)$. Moreover, we have checked for low
values of $k$ that the equation for $\latot$, resulting
from the IOW equations based on \eqref{eqGBi}, are identical
to those based on \eqref{eqGBa}.  Furthermore, in all formulations,
the equations \eqref{sec4eq6ea} and \eqref{sec4eq7ea}
are consistent with \eqref{eqGAa}.

For $k=2,3$, we checked the small $x$ behaviour for $\latot$, 
Eq.\ \eqref{sec4eq1ea}. We again expect the constants $\alpha$
to be the quantum dimensions of the associated conformal field theory. 
Using some results in \cite{dFMS}, these quantum dimensions are 
given by 
\begin{equation} \label{eqGBk}
\alpha_k \eql 1 + 2 \cos \left( \frac{2 \pi}{k+3} \right) \,.
\end{equation}
For $k=2$, the equation for $\latot$ reads (upon taking
$x_\ua=x_\da=x$)
\begin{equation} \label{eqGBl}
(\latot^{\frac{1}{2}}-1)^2 \eql x^2 \latot^\frac{8M+5}{8M+6}
+x \latot^\frac{6M+4}{8M+6} - x \latot^\frac{2M+1}{8M+6} \,,
\end{equation}
which leads to the following small $x$ behaviour
\begin{equation} \label{eqGBm}
\latot \eql 1 + 2 \left( \frac{1+\sqrt{5}}{2} \right) x + o(x^2) \,,
\end{equation}
in agreement with $\alpha_2 = (1+\sqrt{5})/2$ from \eqref{eqGBk};
the extra factor $2$ comes from the sum over the two physical
particles, see Eq.\ \eqref{sec4eq1ea}. For $k=3$ we find
\begin{equation}
(\latot^\frac{1}{2}-1) \eql x \latot^\frac{8M+3}{6(6M+3)}
(\latot^\frac{1}{6}+1)^\frac{1}{3}(\latot^\frac{1}{3}+1)^\frac{2}{3} \,,
\end{equation}
which gives $\alpha_3=2$, consistent with \eqref{eqGBk}.
Note that for the abelian case $k=1$, we find for the small
$x_{\ua,\da}$-behaviour, using \eqref{eqGAd},
\begin{equation} \label{eqGBn}
\latot \eql  1 + (x_\ua+x_\da) + o(x^2) \,,
\end{equation}
in agreement with \eqref{eqGBk} and the fact that for $k=1$
we have an abelian state.

As was the case for the spin polarized states of Sect.\ 6,
also for the non-abelian spin singlet states a particle
behaving as a supercurrent can be identified in the non-magnetic 
limit. The situation here is slightly more complicated than in the case 
of the spin polarized states discussed in Sect.\ 6.
This is because in the formulation above, there is no 
candidate particle with the property that all the statistics 
parameters go to zero in the limit $q \to 0$ (with $q=1/\nu=M+3/2k$).
However, if one acts with $\cC_{2k-1,2k}$ on $\cS_M \bK_e$, 
with $\bK_e$ given by Eq.\ \eqref{eqGBg}, one introduces
a composite with charge $-2k$ and spin $0$, which
has the desired properties. In the $\phi$-sector, the
particle content is changed to one quasihole and 
$2k$ pseudoparticles, of which a few carry spin. 

The possibility to introduce a composite with the right properties
enables one to repeat the discussion of Sect.\ 6, with the only
difference that the flux quantum in this case equals $h/2ke$.
So, also in this case, we can identify a supercurrent in the
non-magnetic limit.

\section{Discussion}

In this paper we derived the $K$-matrix structure for two classes of
so called non-abelian quantum Hall states, putting the results of
\cite{ABGS} on a firmer basis. In doing so, we extensively made
use of a duality between the edge electron and quasihole
excitations. The abelian formalism was extended to include electron 
spin, in order to be able to treat spin singlet states. 
Moreover, we showed that many results
of the abelian $K$-matrix formulation for hierarchy states also
hold for our generalized $K$-matrices, thereby justifying their
name. We would like to stress that the non-abelian states
of \cite{RR,AS} are not hierarchical states; the $K$-matrix 
structure is necessary as a bookkeeping device for the non-abelian
statistics. 

An important concept we did not discuss is the torus degeneracy 
\cite{Wena}; it is not clear at the moment how 
to generalize this to the non-abelian case (some remarks 
are made in Appendix D). 
Another important issue to be settled has to do with the cases
where the pseudoparticles do carry spin (or charge). These may
arise by creating extra composites in the electron sector; by 
the duality, the $\phi$ sector changes accordingly, and 
pseudoparticles carrying spin may arise. The formulas
Eq.\ \eqref{sec2eq2ea} then need a proper adjustment, because
they do not give the same result any more, and the physical 
quantities like the filling factors need to be invariant
under the introduction of extra composites. We would like to
remark that a description in which the pseudoparticles do not
carry spin or charge is possible in the cases we examined,
and the various physical quantities were obtained correctly.

As for the Laughlin wave functions, one would like to have 
a Landau-Ginzburg field theory describing the 
excitations for the non-abelian states. The backbone of 
such a theory will be a Chern-Simons term, in which the gauge 
fields are coupled in a special way. We expect that the 
$K$-matrices derived in this paper will play a crucial role. 
{}From a Landau Ginzburg theory (using the $K$-matrices etc. 
from the electronic sector), one should be able to identify
the possible excitations in the $\phi$-sector, as vortex
solutions of the classical equations of motion. Identifying
this Landau-Ginzburg theory is left for future investigations
(see \cite{LG} for related studies).

Another interesting issue for the non-abelian states is the
determination of the degeneracies of the states when extra
flux is applied through the sample. These degeneracies can
be calculated using conformal field theory techniques,
and can, interestingly, be simulated on a computer using a
special, ultra local, interaction for the electron interaction.
For the Pfaffian, exact counting results were 
obtained in \cite{RRb}; the more general Read-Rezayi states
were treated in \cite{GuR}. Counting results for the  
NASS states will be given elsewhere \cite{ARRS}. 

Finally, while our discussion of  fqH-bases of  
conformal field theories based on quasiparticles with a 
statistics matrix $\bK\oplus\bK^{-1}$ 
was restricted to $(\wh{\mathfrak{sl}_n})_k$ (for $n=2,3$), 
it is obvious that such a description generalizes to more general 
conformal field theories (see \cite{Bo2} for more examples),
even though these may not have an interpretation in the context
of the fractional quantum Hall effect.

\section*{Acknowledgments}

We would like to thank Sathya Guruswamy, Nick Read and 
Ole Warnaar for useful discussions.  
This research was financially supported in part by the foundation FOM of 
the Netherlands and the Australian Research Council.

\appendix

\section{Basic hypergeometric series}

Consider the basic hypergeometric series
\begin{multline}
{_r}\phi_s(a_1,\ldots,a_r;b_1,\ldots,b_s;q,z) \eql \\ 
\sum_{m\geq0} \frac{(a_1;q)_m(a_2;q)_m \ldots (a_r;q)_m}{(q;q)_m 
(b_1;q)_m \ldots
 (b_s;q)_m} \left( (-1)^m q^{\frac{1}{2} m(m-1)}\right)^{1+s-r} z^m \,,
\end{multline}
where 
\begin{equation}
(a;q)_n \eql \prod_{k=0}^{n-1} (1-aq^k) \,.
\end{equation}
We have the $q$-Pfaff-Saalsch\"utz sum \cite{GR,Sla}
\begin{equation} \label{eqUCAa}
{_3}\phi_2(a,b,q^{-n};c,abq^{1-n}/c;q,q)\eql
\frac{(c/a;q)_n (c/b;q)_n}{(c;q)_n(c/ab;q)_n} \,,
\end{equation}
Taking $b=0$ in \eqref{eqUCAa} gives the $q$-Chu-Vandermonde sum
\begin{equation} \label{eqUCAb}
{_2}\phi_1(a,q^{-n};c;q,q)\eql
\frac{(c/a;q)_n}{(c;q)_n} a^n \,.
\end{equation}

Now, taking $a=q^{-m_1}$, $b=q^{M_1+M_2-(m_1+m_1)+1}$, 
$c=q^{M_2-(m_1+m_2)+1}$ and $n=m_2$ in \eqref{eqUCAa} gives 
\begin{multline}
\qbin{M_1}{m_1} \qbin{M_2}{m_2} \eql  \\
\sum_{m\geq0} q^{(m_1-m)(m_2-m)}
\qbin{M_1-m_2}{m_1-m} \qbin{M_2-m_1}{m_2-m} \qbin{M_1+M_2-(m_1+m_2)+m}{m}\,.
\end{multline}
Taking $a=q^{-m_1}$,  
$c=q^{M_2-(m_1+m_2)+1}$ and $n=m_2$ in \eqref{eqUCAb} gives 
\begin{equation}
\frac{1}{\qn{m_1}} \ \qbin{M_2}{m_2} \eql
\sum_{m\geq0} q^{(m_1-m)(m_2-m)} \frac{1}{\qn{m}\qn{m_1-m}} 
\qbin{M_2-m_1}{m_2-m} \,,
\end{equation}
while taking $a=q^{-m_1}$, $n=m_2$  and $c=0$ in \eqref{eqUCAb}
gives
\begin{equation}
\frac{1}{\qn{m_1}\qn{m_2}} \eql
\sum_{m\geq0} q^{(m_1-m)(m_2-m)} \frac{1}{\qn{m}\qn{m_1-m} \qn{m_2-m}}\,.
\end{equation}

\section{The principal subspace}

In this appendix we review an 
important result of \cite{FS,Geo} which is used 
throughout the paper.  Consider an affine Lie algebra $\wh{\mathfrak g}_k$
(see, e.g., \cite{Kac} for notation and definitions).
If $L(\Lambda)$ is the integrable highest weight
module of $\wh{\mathfrak g}_k$ with highest weight $\Lambda$ and 
highest weight vector $v_{\Lambda}$, then the principal subspace
$W(\Lambda)\subset L(\Lambda)$ is defined to be the subspace 
generated from $v_\Lambda$ by the negative modes of the positive simple 
root currents $J_{\alpha_i}(z)$.

The character of the principal subspace $W(\Lambda)$ of the integrable 
highest weight module $L(\Lambda)$ for $\Lambda=k_0\Lambda_0+k_j\Lambda_j$
($1\leq j\leq n$, $k_0+k_j=k$) of
$(\wh{\mathfrak{sl}_{n+1}})_k$ was determined in \cite{FS,Geo}.%
\footnote{The character of the principal 
subspace $W(\Lambda)$ for more general level $k$ modules
$L(\Lambda)$ is apparently not yet known.}
It is given by the UCPF 
\begin{equation} \label{eqAPBa}
\charac_W \eql \sum_{\bp} \ \left( \prod z_i^{sp_i^{(s)}} \right) \,
\frac{q^{ \frac{1}{2} \bp\cdot \bK \cdot \bp
+ \bQ_j\cdot \bp} }{\prod_i \prod_s \qn{p_i^{(s)}}} \,,
\end{equation}
where 
\begin{equation} \label{eqAPBb}
\bK \eql \bA_n \otimes \bB_k \,, \qquad
\bQ_j \eql \bfe_j \otimes (\underbrace{0,\ldots,0}_{k_0},1,2,\ldots,k_j)\,,
\end{equation}
and $z_i$ denotes the (generalized) fugacity of the current $J_{\alpha_i}$.
Also, $(\bA_n)_{ij} = 2\delta_{ij} - \delta_{i-1,j} - \delta_{i+1,j}$
is the Cartan matrix of $\mathfrak{sl}_{n+1}$, i.e.\
\begin{equation} \label{eqAPBc}
\bA_n \eql \begin{pmatrix}
2 & -1 & \\
-1 & 2 & -1 & \\
 & -1 & 2 & -1 & \\
&& \ddots & \ddots & \ddots & \\
&&& -1 & 2 & -1 \\
&&& & -1 & 2 \end{pmatrix} \,,
\end{equation}
and $(\bB_k)_{rs} \eql \min(r,s)_{\lvert r,s=1,\ldots,k}$, i.e.\
\begin{equation}  \label{eqAPBd}
\bB_k \eql \begin{pmatrix}
1 & 1 & 1 & \ldots & 1 \\
1 & 2 & 2 & \ldots & 2 \\
1 & 2 & 3 & \ldots & 3 \\
\vdots & & & \ddots & \vdots \\
1 & 2 & 3 & \ldots & k \end{pmatrix} \,.
\end{equation}
Furthermore, 
in \eqref{eqAPBa}, we have written $\bp= (p_j^{(s)})_{j=1,\ldots,n}^{
s=1,\ldots,k}$ with respect to $(\bA_n)_{ij} \otimes (\bB_k)_{rs}$.

\section{$(\wh{\mathfrak{sl}_2})_{k,M}$ character}

The UCPF character for $(\wh{\mathfrak{sl}_2})_{k=1,M}$ was discussed
in \cite{ES} (see also \cite{BM}).  Here we discuss the $q$-Pfaffian case,
i.e.\ $k=2$.  For convenience we put $q=M+1$.

\subsection{Quasihole sector}

In \cite{Scb}, finitized partition sums 
$X_L = X_{l=\frac{8L-q-2}{16q}}$ and $Y_L = Y_{l=\frac{8L+q-6}{16q}}$ 
for the quasihole sector of the $q$-pfaffian CFT were introduced.
$X_L$ ($Y_L$) are restricted by requiring that the total charge
be an even (odd) multiple of $\frac{1}{2q}$. In \cite{Scb}, it
was established that the following recursion relations hold
\begin{eqnarray} \label{eqAPCaa}
X_L & \eql & X_{L-2q} + x q^{\frac{8L-q-2}{16q}} (Y_{L} + Y_{L-q})\,, 
  \nonumber\\
Y_L & \eql & Y_{L-2q} + x q^{\frac{8L+q-6}{16q}} X_{L-1} \,,
\end{eqnarray}
or, equivalently,
\begin{equation} \label{eqAPCab}
X_L \eql X_{L-2q} + q^{\frac{1}{2}} \left( X_{L-q} - X_{L-3q} \right)
 + x^2 q^{\frac{2L-1}{2q}} X_{L-1} \,.
\end{equation}
By putting $X_L/X_{L-q} \sim \latot^{1/2q}$, for large $L$, we 
reproduce equation \eqref{eqFAba}.
To build the entire spectrum of the $(\wh{\mathfrak{sl}_2})_{k=2,M}$ 
conformal field theory we need $3q$ sectors whose initial conditions are
given in Table \ref{tabAPCa}.  The vacua of the 
sectors are labeled by, respectively, charge and the $\mathfrak{sl}_2$
irrep in which they appear for $M=0$ (the labels $\id$, $\sigma$ and 
$\psi$ stand for the $\mathfrak{sl}_2$ singlet, doublet and triplet,
respectively, in analogy with the Ising model).  The parameter $r$
takes the values $r=0,1,\ldots,q-1$.

\begin{table}[ht]
\begin{tabular}{|l|l|c|} \hline
sector & initial conditions & $\bQ_\phi$ \\ \hline
$\mid-\frac{re}{q},\id\rangle$ & $X_{s}=1, Y_{s}=0$ & 
$(0,\txt{\frac{s}{2q}})$ \\
$\mid-\frac{(2r+1)e}{2q},\sigma\rangle$ & 
  $X_{2q-r}=xq^{\frac{15q-2-8r}{16q}}, 
Y_{2q-r}=1$ 
& $(-\txt{\frac{1}{2}},\txt{\frac{5q-1-2r}{4q}})$ \\
$\mid-\frac{re}{q},\psi\rangle$ & $X_{s}=1, Y_{s}=0$&
$(0,\txt{\frac{s}{2q}})$ \\ \hline
\end{tabular}
\caption{} \label{tabAPCa}
\end{table}

The solutions to \eqref{eqAPCaa} can be written in terms of finitized
UCPFs with (cf.\ \eqref{eqFAb})
\begin{equation} \label{eqAPCba}
\bK_\phi \eql \begin{pmatrix}
1 & -\frac{1}{2} \\ -\frac{1}{2} & \frac{q+1}{4q} \end{pmatrix} \,.
\end{equation}
Indeed, the recursion relations \eqref{eqEBb}, with $\bK=\bK_\phi$
and $\bQ+\bu=0$, are explicitly given by
\begin{eqnarray}\label{eqAPCbb}
P_{L_1,L_2} & \eql & P_{L_1-1,L_2} + q^{L_1-\frac{1}{2}} 
  P_{L_1-1,L_2+\frac{1}{2}} \,, \nonumber\\
P_{L_1,L_2} & \eql & P_{L_1,L_2-1} + xq^{L_1-\frac{q+1}{8q}} 
  P_{L_1+\frac{1}{2},L_2-\frac{q+1}{4q}} \,, 
\end{eqnarray}
and lead to \eqref{eqAPCaa} upon identifying
\begin{equation}\label{eqAPCbc}
X_L \eql q^{ \frac{1}{4} Q_1^2} P_{0,\frac{L}{2q}}\,,\qquad
Y_L \eql q^{\frac{1}{4} Q_1^2 -\frac{1}{16}} P_{-\frac{1}{2},\frac{L}{2q}+ 
  \frac{q-1}{4q}} \,.
\end{equation}
The values for $\bQ_\phi$ in each sector are listed in 
Table \ref{tabAPCa}, while the parameters 
$s=0,\ldots,2q-1$, in Table \ref{tabAPCa}, are given 
in Table \ref{tabAPCb}.

\begin{table}[ht]
\begin{tabular}{|c|c|c|c|c|c|c|c|c|c|} \hline
sector&$\mid0,\id\rangle$ & $\mid-\frac{e}{q},\psi\rangle$ &
$\mid -\frac{2e}{q},\id\rangle$ & $\ldots$ & $\mid0,\psi\rangle$ & 
$\mid-\frac{e}{q},\id\rangle$ & $\mid -\frac{2e}{q},\psi\rangle$ & $\ldots$ & 
\\ \hline 
$s$   & $0$ & $1$ & $2$ & 
$\ldots$ & $q$ & $q+1$ & $q+2$ & $\ldots$ & $2q-1$ \\ \hline
\end{tabular}
\caption{} \label{tabAPCb}
\end{table}

\subsection{Electron sector}

For the electron sector of the $q$-pfaffian, the paper
\cite{Scb} introduced the truncated partition sums $\Omega_L$,
which contain all states constructed from the 
edge electron operator $\Psi_{-s}$ with $s\leq L-\frac{q-1}{2}$.
It satisfies the recursion relation
\begin{equation} \label{eqAPCc}
\Omega_L \eql \Omega_{L-1} + y q^{L-\frac{1}{2}(q-1)} \Omega_{L-q} +
y^2 q^{2L-(2q-1)} \Omega_{L-2q} - y^3 q^{3L- \frac{1}{2} (9q-5)} 
\Omega_{L-3q}\,,
\end{equation}
and results in Eq.\ \eqref{eqFAd} by putting $\Omega_L/\Omega_{L-1}\sim
\mutot$ for large $L$.
In this case the recursion relation does not appear to be solved 
by finitized UCPFs.  However, there exists a recursion relation, 
leading to the same equation for $\mutot$, that is solved by a finitized
UCPF and differs from the solution to \eqref{eqAPCc} by terms of 
order $q^L$ (i.e.\ belongs to the same universality class, 
see the discussion in Sect.\ 5.3) and thus gives 
the correct solution in the limit $L\to\infty$.
The UCPF is based on the $K$-matrix (cf.\ \eqref{eqFAc})
\begin{equation} \label{eqAPCd}
\bK_e \eql \begin{pmatrix}
q+1 & 2q \\ 2q & 4q \end{pmatrix} \,.
\end{equation}
The initial conditions and values for $\bQ_e$ in each sector are listed 
in Table \ref{tabAPCc}.

\begin{table}[ht]
\begin{tabular}{|l|l|c|} \hline
sector & initial conditions & $\bQ_e$  \\ \hline 
$\mid -\frac{re}{q} ,\id\rangle$ & $\Omega_{r-1}=\ldots=\Omega_{q+r-1}=1$  & 
  $(r,2r)$ \\
$\mid-\frac{(2r+1)}{2q},\sigma\rangle$ &$\Omega_{r}=\ldots=\Omega_{q+r-1}=1$
  & $(r,2r+1)$ \\
$\mid-\frac{re}{q},\psi\rangle$ & $\Omega_{-q+r-1}= q^{\frac{1}{2}(q-1)-r}/y,
\Omega_r=1$ &  $(r-1,2r+1;-1,-1,-1,\ldots)$  \\ \hline 
\end{tabular}
\caption{} \label{tabAPCc}
\end{table}

There is a slight subtlety in the case of the sectors $|-re/q,\psi\rangle$.
These vectors do not correspond to an extremal vector in the 
$(\wh{\mathfrak{sl}_2})_{k=2,M}$ modules.  Thus the results of App.\ B do not
apply.  While the exclusion statistics of the currents is unchanged,
and hence the $K$-matrix is still given by \eqref{eqAPCd}, it can easily 
be shown that the extremal vectors in the modules cannot be reproduced 
by any two dimensional vector $\bQ$.  In fact, to correctly reproduce 
the extremal vectors one needs an infinite dimensional vector $\bQ$
(given in Table \ref{tabAPCc}) with a corresponding infinite 
dimensional $K$-matrix $\bK_e^{(\infty)}$ 
that is equivalent to \eqref{eqAPCd} by the
composite construction.  Specifically, one introduces derived matrices
$\bK_e^{(n)}$
and associated generalized fugacities $\bz^{(n)}$ by
\begin{equation} \label{eqAPCe}
\bK_e^{(1)} \eql \cC_{12} \bK_e \eql \begin{pmatrix}
q+1 & 2q+1 & 3q+1 \\ 
2q+1 & 4q & 6q \\
3q+1 & 6q & 9q+1 \end{pmatrix} \,, \qquad \bz^{(1)} \eql \begin{pmatrix}
z \\ z^2 \\ z^3 \end{pmatrix} \,,
\end{equation}
\begin{equation} \label{eqAPCea}
\bK_e^{(2)} \eql \cC_{23}\bK_e^{(1)} \eql \begin{pmatrix}  
q+1 & 2q+1 & 3q+1 & 5q+2\\ 
2q+1 & 4q & 6q+1 & 10q \\
3q+1 & 6q+1 & 9q+1 & 15q+1 \\
5q+2 & 10q & 15q+1 & 25q+1 
\end{pmatrix} \,, \qquad
\bz^{(2)} \eql \begin{pmatrix}
z \\ z^2 \\ z^3 \\ z^5 \end{pmatrix} \,,
\end{equation}
and, ultimately,
\begin{multline} \label{eqAPCf}
\bK_e^{(\infty)} \eql \lim_{n\to\infty} \bK_e^{(n)} \eql \lim_{n\to\infty}\  
\cC_{2,n}\cC_{2,n-1}\ldots\cC_{23}\cC_{12} \bK_e \\ \eql
\begin{pmatrix}
q+1  & 2q+1 & 3q+1  & 5q+2  & 7q+3 & \ldots & (2k+1)q+k  & \ldots \\ 
2q+1 & 4q   & 6q+1  & 10q+1 & 14q+1& \ldots & 2(2k+1)q+1 & \ldots \\
3q+1 & 6q+1 & 9q+1  & 15q+1 & 21q+2& \ldots &            &  \\
5q+2 & 10q+1& 15q+1 & 25q+1 & 35q+1& \ldots &            &  \\
7q+3 & 14q+1& 21q+2 & 35q+1 & 49q+1& \ldots &            &  \\
\vdots&\vdots&\vdots&\vdots &\vdots& \ddots & \vdots     &  \\
     &      &       &       &      &        &(2k+1)^2q+1 &  \\
     &      &       &       &      &        &            & \ddots 
\end{pmatrix} \,,
\end{multline}
while
\begin{equation}\label{eqAPCfa}
\bz^{(\infty)} \eql (z,z^2;z^3,z^5,z^7,\ldots)\,.
\end{equation}
For every finite $n$, the UCPF based on $(\bK_e^{(n)};\bQ_e^{(n)})$,
where $\bQ_e^{(n)}$ is the $(n+2)$-dimensional truncation of the vector 
$\bQ_e$ in Table \ref{tabAPCc}, gives an accurate description of 
the module up to some level (which appears to be at at least polynomially 
increasing with $n$).  To describe the entire module accurately, one needs
to take the limit $n\to\infty$. 

\subsection{The character}

Combining the $3q$ sectors in Tables \ref{tabAPCa} and \ref{tabAPCb}
should reproduce the spectrum of the chiral $(\wh{\mathfrak{sl}_2})_{2,M}$
conformal field theory.  Consider the combination of UCPFs
\begin{equation} \label{eqAPCg}
Z_{\text{tot}} \eql \sum_{k=1}^{3q} \ a_{(k)} Z_\infty(\bK_e;\bQ^{(k)}_e)
  Z_{\infty/2}(\bK_\phi;\bQ^{(k)}_\phi,\bu^{(k)}_\phi) \,,
\end{equation}
where the coefficients $a_{(k)}$ and vectors $\bQ_e^{(k)}$, $\bQ^{(k)}_\phi =
- \bu^{(k)}_\phi$ are given in Table \ref{tabAPCd}, and where
\begin{multline}
Z_{\infty/2}(\bK_\phi;\bQ^{(k)}_\phi,\bu^{(k)}_\phi)  \\  \equiv 
q^{\frac{1}{4}(Q_1^{(k)})^2} 
\left( Z(\bK_\phi;\bQ^{(k)}_\phi,\bu^{(k)}_\phi) +
q^{-\frac{1}{16}} Z(\bK_\phi;\bQ^{(k)}_\phi,\bu^{(k)}_\phi - 
\begin{pmatrix} 1/2 \\ 0 \end{pmatrix} ) \right) \,,
\end{multline}
corresponds to the limit
\begin{equation}
\lim_{L\to \infty} \sum_{t=0}^{2q-1} \left( X_{L-t} + Y_{L-t} \right) \,.
\end{equation}
\begin{table}[ht]
\begin{tabular}{|l|c|c|c|} \hline
sector & $\bQ_\phi$ & $\bQ_e$ & $a_{(k)}$ \\ \hline
$\mid-\frac{re}{q},\id\rangle$ & 
$(0,\txt{\frac{s}{2q}})$ & $(r,2r)$ & $x^{-2r} q^{\frac{r^2}{2q}}$ \\
$\mid-\frac{(2r+1)e}{2q},\sigma\rangle$ & 
   $(-\txt{\frac{1}{2}},\txt{\frac{5q-1-2r}{4q}})$ & $(r,2r+1)$ & 
 $x^{-(2r+1)} q^{\frac{(2r+1)^2}{8q} + \frac{1}{16}}$ \\
$\mid-\frac{re}{q},\psi\rangle$ &
$(0,\txt{\frac{s}{2q}})$ & $(r-1,2r+1;-1,-1,\ldots)$ & 
  $x^{-2r} q^{\frac{r^2}{2q}+ \frac{1}{2}}$ \\ \hline
\end{tabular}
\caption{} \label{tabAPCd}
\end{table}

We have numerically checked that \eqref{eqAPCg} indeed equals 
the $(\wh{\mathfrak{sl}_2})_{k=2,M}$ character 
\begin{equation} \label{eqAPCh}
Z_{\text{tot}} \eql
\frac{1}{\qn{\infty}} \sum_{n\in\ZZ}  \left( x^{2n} 
  q^{\frac{1}{2q}n^2} \prod_{k\geq1} (1+q^{k-\frac{1}{2}}) + 
 x^{2n+1} q^{ \frac{1}{2q}(n+\frac{1}{2})^2 +\frac{1}{16}} 
\prod_{k\geq1}(1+q^{k}) \right) \,,
\end{equation}
corresponding to a free fermion and a boson compactified on a circle of 
radius $R^2=q$.
It should be possible to prove the equality of \eqref{eqAPCg} and 
\eqref{eqAPCh} along the lines of \cite{Bo2} (see also App.\ D).
Finally, we note that the number of summands in \eqref{eqAPCg}
equals the torus degeneracy for the $q$-Pfaffian computed in 
\cite{GWW}.

\section{$(\wh{\mathfrak{sl}_3})_{k,M}$ character}

We will restrict the discussion in this section to 
$(\wh{\mathfrak{sl}_3})_{k,M}$ for level $k=1$.

\subsection{Quasihole sector}

The recursion relation for the quasiholes $(\phi_\ua,\phi_\da)$
in $(\wh{\mathfrak{sl}_3})_{k,M}$ for $k=1$ and  $M=0$ 
was worked out in \cite{Sca,BSb}.  
The generalization to arbitrary $M$ reads
\begin{equation} \label{eqAPDa}
X_L \eql X_{L-(2M+3)} + (x_\ua+x_\da) 
  q^{ \frac{2L-(M+2)}{2(2M+3)}} X_{L-(M+2)} +
  x_\ua x_\da q^{\frac{2L-1}{2M+3} }X_{L-1} \,.
\end{equation}
By putting $X_L/X_{L-1} \sim \latot^{\frac{1}{2M+3}}$ 
we recover the IOW-equation \eqref{eqGAd}.
To build the entire spectrum out of quasiholes and electrons we 
need $3M+4$ sectors 
whose initial conditions are given in Table \ref{tabAPDa}.
The vacua of the sectors are labeled by, respectively, charge, spin,
and the $\mathfrak{sl}_3$ irrep in which they occur for $M=0$.
The parameter $r$ takes the values $r=1,2,\ldots,M+1$.

The solution to \eqref{eqAPDa} can be written in terms of 
finitized UCPFs (see \eqref{eqEBa}).  Indeed, the recursion relations 
\eqref{eqEBb} with (see \eqref{eqGAc})
\begin{equation}\label{eqAPDb}
\bK_\phi \eql \txt{\frac{1}{2M+3}} \begin{pmatrix}
M+2 & -(M+1) \\ -(M+1) & M+2 \end{pmatrix} \,,
\end{equation}
and $\bQ+\bu=(0,0)$ are explicitly given by
\begin{eqnarray}\label{eqAPDc}
P_{L_1,L_2} & \eql & P_{L_1-1,L_2} + x_\ua q^{L_1-\frac{M+2}{2(2M+3)}}   
   P_{L_1-\frac{M+2}{2M+3},L_2+\frac{M+1}{2M+3}} \nonumber\\
P_{L_1,L_2} & \eql & P_{L_1,L_2-1} + x_\da q^{L_2-\frac{M+2}{2(2M+3)}}   
   P_{L_1+\frac{M+1}{2M+3},L_2-\frac{M+2}{2M+3}} 
\end{eqnarray}
Setting $X_L\equiv P_{L/(2M+3),L/(2M+3)}$ leads to \eqref{eqAPDa}.
The values for $\bQ=-\bu$ in the various sectors, as determined by the 
initial conditions, are given in Table \ref{tabAPDa}.
\bigskip

\begin{table}[ht]
\begin{tabular}{|l|l|l|c|c|} \hline
sector & initial conditions & 
  $\bQ_\phi$  \\ \hline 
$\mid 0,-,\id\rangle$ & $X_0=1$ & 
  $(0,0)$ \\
$\mid -\frac{2re}{3},-,{\mathbf 3}\rangle$ &   
 $X_{2M+3-r}=1$  & $(\txt{\frac{2M+3-r}{2M+3}},
  \txt{\frac{2M+3-r}{2M+3}} )$ \\
$\mid -\frac{(2r-1)e}{3},\ua,{\mathbf 3}^*\rangle$ &$X_{M+2-r}=1$& 
 $(\txt{\frac{M+2-r}{2M+3}},\txt{\frac{M+2-r}{2M+3}} )$ \\
$\mid -\frac{(2r-1)e}{3},\da,{\mathbf 3}^*\rangle$ &$X_{M+2-r}=1$&
$(\txt{\frac{M+2-r}{2M+3}},\txt{\frac{M+2-r}{2M+3}} )$ \\ \hline 
\end{tabular}
\caption{}  \label{tabAPDa}
\end{table}

\subsection{Electron sector}

The recursion relations for the electrons $(\Psi_\ua,\Psi_\da)$ 
are given by 
\begin{equation}\label{eqAPDd}
\Omega_L \eql \Omega_{L-1} + (y_\ua+y_\da) 
q^{L-\frac{M}{2}} \Omega_{L-(M+2)} + y_\ua y_\da 
q^{2L-(2M+1)} \Omega_{L-(2M+3)} \,.
\end{equation} 
with initial conditions listed in Table \ref{tabAPDb}.
They can be solved by finitized UCPFs with (see \eqref{eqGAl})
\begin{equation}\label{eqAPDe}
\bK_e \eql \begin{pmatrix}
M+2 & M+1 \\ M+1 & M+2 \end{pmatrix}\,,
\end{equation}
and $\bQ+\bu=(1,1)$, by putting $\Omega_L = P_{L,L}$. The values
for $\bQ_e$ in the various sectors are listed in Table \ref{tabAPDb}.

\begin{table}[ht]
\begin{tabular}{|l|l|l|c|c|} \hline
sector & initial conditions & $\bQ_e$ \\ \hline 
$\mid 0,-,\id\rangle$ & $\Omega_{-1}=\Omega_0=\ldots=\Omega_{M}=1$ 
  &  $(0,0)$ \\
$\mid -\frac{2re}{3},-,{\mathbf 3}\rangle$ &   
 $\Omega_{r-1}=\Omega_r=\ldots=\Omega_{M+r}=1$ & $(r,r)$ \\
$\mid -\frac{(2r-1)e}{3},\ua,{\mathbf 3}^*\rangle$ &
 $\Omega_{r-1}=\Omega_r=\ldots=\Omega_{M+r-1}=1$   & $(r-1,r)$ \\
  &  $\Omega_{M+r}=1+y_\ua q^{\frac{1}{2}(M+2)+(r-1)}$ & \\
$\mid -\frac{(2r-1)e}{3},\da,{\mathbf 3}^*\rangle$ &
  $\Omega_{r-1}=\Omega_r=\ldots=\Omega_{M+r}=1$  & $(r,r)$ \\ \hline 
\end{tabular}
\caption{}  \label{tabAPDb}
\end{table}

\subsection{Character}

Combining the $3M+4$ sectors in Tables \ref{tabAPDa} and 
\ref{tabAPDb} should reproduce 
the spectrum of the chiral $(\wh{\mathfrak{sl}_3})_{k=1,M}$ conformal 
field theory.
Indeed, consider the following combination of UCPFs 
\begin{equation}\label{eqAPDf}
Z_{\text{tot}} \eql \sum_{k=0}^{3M+3} \ a_{(k)} 
Z_\infty(\bK_e;\bQ_e^{(k)}) Z_\infty(\bK_\phi;\bQ_\phi^{(k)}) \,,
\end{equation}
where the coefficients $a_{(k)}$ are defined in Table \ref{tabAPDc}.

\begin{table}[ht]
\begin{tabular}{|l|l|l|c|c|} \hline
sector & $\bQ_\phi$ & $\bQ_e$ & $a_{(k)}$ \\ \hline 
$\mid 0,-,\id\rangle$ & $(0,0)$ &  $(0,0)$ & $1$ \\
$\mid -\frac{2re}{3},-,{\mathbf 3}\rangle$ & 
$(\txt{\frac{2M+3-r}{2M+3}},  \txt{\frac{2M+3-r}{2M+3}} )$   
 & $(r,r)$ & $(x_\ua x_\da)^{-r} q^{\frac{r^2}{2M+3}}$ \\
$\mid -\frac{(2r-1)e}{3},\ua,{\mathbf 3}^*\rangle$ &
  $(\txt{\frac{M+2-r}{2M+3}},\txt{\frac{M+2-r}{2M+3}} )$
 & $(r-1,r)$ &  $x_\ua (x_\ua x_\da)^{-r} q^{\frac{2r(r-1)+M+2}{2(2M+3)}}$  \\
$\mid -\frac{(2r-1)e}{3},\da,{\mathbf 3}^*\rangle$ &
  $(\txt{\frac{M+2-r}{2M+3}},\txt{\frac{M+2-r}{2M+3}} )$
   & $(r,r)$ &  
  $x_\da (x_\ua x_\da)^{-r} q^{\frac{2r(r-1)+M+2}{2(2M+3)}}$\\ \hline 
\end{tabular}
\caption{}  \label{tabAPDc}
\end{table}

We claim that \eqref{eqAPDf} 
equals the $(\wh{\mathfrak{sl}_3})_{k=1,M}$ character
\begin{equation} \label{eqAPDg}
Z_{\text{tot}} \eql \frac{1}{\qn{\infty}^2} \sum_{p_i\in\ZZ}
(x_\ua^{p_1} x_\da^{p_2}) q^{\frac{1}{2} \bp\cdot \bK_\phi \cdot \bp}
\end{equation}
corresponding to the 
partition function of two chiral bosons on the deformed weight lattice 
of $\mathfrak{sl}_3$.  E.g., \eqref{eqAPDg} for $M=0$ is 
precisely the Frenkel-Kac character (see, e.g., \cite{Kac})
of the sum of the integrable 
highest weight modules of $(\wh{\mathfrak{sl}_3})$ at level $k=1$.

To prove this claim, first observe that we can rewrite \eqref{eqAPDf}
as a sum over $2M+3$ sectors by using \eqref{eqEBad}.  
Specifically,
\begin{equation}
Z_\infty(\bK_\phi, \begin{pmatrix} \frac{2M+3-r}{2M+3} \\ \frac{2M+3-r}{2M+3}
\end{pmatrix} ) + x_\da q^{ \frac{M+2-2r}{2(2M+3)}} Z_\infty(\bK_\phi, 
\begin{pmatrix} \frac{M+2-r}{2M+3} \\ \frac{M+2-r}{2M+3} 
\end{pmatrix} ) \eql Z_\infty(\bK_\phi, 
\begin{pmatrix} \frac{2M+3-r}{2M+3} \\ \frac{-r}{2M+3}
\end{pmatrix} ) \,,
\end{equation}
after which the claim follows by applying the statements of Theorem 4.1 
and Corollary 5.2 in \cite{Bo2}.

Note that even though we use $3M+4$ sectors in generating the 
entire spectrum from
the recursion relations \eqref{eqAPDa} and \eqref{eqAPDd}, the
$(\wh{\mathfrak{sl}_3})_{k=1,M}$
partition function \eqref{eqAPDg} can be written in terms of UCPFs
based on $\bK=\bK_e\oplus \bK_\phi$ using only $2M+3$ sectors.
So, even though the UCPF form of a partition function is not unique, 
and we do not understand the precise relation between the number of sectors
and the torus degeneracy in the sense of Wen et al.\ \cite{Wena},
it is satisfying to see that the number $2M+3$ equals $\det\bK_e$
which is the torus degeneracy for abelian fqH systems.




\end{document}